\documentclass[twocolumn]{aastex63}
\usepackage{amsmath,amstext}
\usepackage{mathrsfs}
\usepackage{subfigure}
\usepackage{tipa}
\usepackage[utf8]{inputenc} 
\usepackage[T1]{fontenc} 
\usepackage{float}

\newcommand\wobble{\texttt{wobble}~}

\newcommand\radvel{\texttt{radvel~}}

\newcommand\mps{m\,s$^{-1}$~}
\newcommand\mpsp{m\,s$^{-1}$}
\newcommand\shk{$S_{HK}$~}
\newcommand\shkp{$S_{HK}$}
\newcommand\ha{H$\alpha$~}
\newcommand\hap{H$\alpha$}
\newcommand\iha{$I_{\textrm{H}\alpha}$~}
\newcommand\ihap{$I_{\textrm{H}\alpha}$}
\newcommand{\unit}[1]{\ensuremath{\, \mathrm{#1}}}

\shorttitle{A terrestrial mass planet candidate orbiting GJ 1151}
\shortauthors{Mahadevan et al.}
\graphicspath{{./}{figures/}}

\begin{document}

\title{The Habitable-zone Planet Finder Detects a Terrestrial-mass Planet Candidate Closely Orbiting Gliese 1151: The Likely Source of Coherent Low-frequency Radio Emission from an Inactive Star}

\correspondingauthor{Suvrath Mahadevan}
\email{suvrath@astro.psu.edu}
\author[0000-0001-9596-7983]{Suvrath Mahadevan}
\affil{Department of Astronomy \& Astrophysics, The Pennsylvania State University, 525 Davey Laboratory, University Park, PA, 16802, USA}
\affil{Center for Exoplanets and Habitable Worlds, 525 Davey Laboratory, University Park, PA, 16802, USA}

\author[0000-0001-7409-5688]{Guðmundur Stefánsson}
\altaffiliation{Henry Norris Russell Fellow}
\affiliation{Princeton University, Department of Astrophysical Sciences, 4 Ivy Lane, Princeton, NJ 08540, USA}

\author[0000-0003-0149-9678]{Paul Robertson}
\affil{Department of Physics and Astronomy, The University of California, Irvine, Irvine, CA 92697, USA}

\author[0000-0002-4788-8858]{Ryan C. Terrien}
\affil{Department of Physics and Astronomy, Carleton College, One North College Street, Northfield, MN 55057, USA}

\author[0000-0001-8720-5612]{Joe P.\ Ninan}
\affil{Department of Astronomy \& Astrophysics, The Pennsylvania State University, 525 Davey Laboratory, University Park, PA, 16802, USA}
\affil{Center for Exoplanets and Habitable Worlds, 525 Davey Laboratory, University Park, PA, 16802, USA}

\author[0000-0002-5034-9476]{Rae J. Holcomb}
\affil{Department of Physics and Astronomy, The University of California, Irvine, Irvine, CA 92697, USA}

\author[0000-0003-1312-9391]{Samuel Halverson}
\affil{Jet Propulsion Laboratory, 4800 Oak Grove Drive, Pasadena, CA 91109, USA}

\author[0000-0001-9662-3496]{William D. Cochran}
\affil{McDonald Observatory and Department of Astronomy, The University of Texas at Austin, 2515 Speedway, Austin, TX 78712, USA}
\affil{Center for Planetary Systems Habitability, The University of Texas at Austin, 2515 Speedway, Austin, TX 78712, USA}

\author[0000-0001-8401-4300]{Shubham Kanodia}
\affil{Department of Astronomy \& Astrophysics, The Pennsylvania State University, 525 Davey Laboratory, University Park, PA, 16802, USA}
\affil{Center for Exoplanets and Habitable Worlds, 525 Davey Laboratory, University Park, PA, 16802, USA}

\author[0000-0002-4289-7958]{Lawrence W. Ramsey}
\affil{Department of Astronomy \& Astrophysics, The Pennsylvania State University, 525 Davey Laboratory, University Park, PA, 16802, USA}
\affil{Center for Exoplanets and Habitable Worlds, 525 Davey Laboratory, University Park, PA, 16802, USA}

\author[0000-0003-1915-5670]{Alexander Wolszczan}
\affil{Department of Astronomy \& Astrophysics, The Pennsylvania State University, 525 Davey Laboratory, University Park, PA, 16802, USA}
\affil{Center for Exoplanets and Habitable Worlds, 525 Davey Laboratory, University Park, PA, 16802, USA}

\author[0000-0002-7714-6310]{Michael Endl}
\affil{McDonald Observatory and Department of Astronomy, The University of Texas at Austin, 2515 Speedway, Austin, TX 78712, USA}
\affil{Center for Planetary Systems Habitability, The University of Texas at Austin, 2515 Speedway, Austin, TX 78712, USA}

\author[0000-0003-4384-7220]{Chad F.\ Bender}
\affil{Steward Observatory, The University of Arizona, 933 N.\ Cherry Ave, Tucson, AZ 85721, USA}

\author[0000-0002-2144-0764]{Scott A. Diddams}
\affil{Time and Frequency Division, National Institute of Standards and Technology, 325 Broadway, Boulder, CO 80305, USA}
\affil{Department of Physics, University of Colorado, 2000 Colorado Avenue, Boulder, CO 80309, USA}

\author[0000-0002-0560-1433]{Connor Fredrick}
\affil{Time and Frequency Division, National Institute of Standards and Technology, 325 Broadway, Boulder, CO 80305, USA}
\affil{Department of Physics, University of Colorado, 2000 Colorado Avenue, Boulder, CO 80309, USA}

\author[0000-0002-1664-3102]{Fred Hearty}
\affil{Department of Astronomy \& Astrophysics, The Pennsylvania State University, 525 Davey Laboratory, University Park, PA, 16802, USA}
\affil{Center for Exoplanets and Habitable Worlds, 525 Davey Laboratory, University Park, PA, 16802, USA}

\author[0000-0002-0048-2586]{Andrew Monson}
\affil{Department of Astronomy \& Astrophysics, The Pennsylvania State University, 525 Davey Laboratory, University Park, PA, 16802, USA}
\affil{Center for Exoplanets and Habitable Worlds, 525 Davey Laboratory, University Park, PA, 16802, USA}

\author[0000-0001-5000-1018]{Andrew J. Metcalf}
\affiliation{Space Vehicles Directorate, Air Force Research Laboratory, 3550 Aberdeen Ave. SE, Kirtland AFB, NM 87117, USA}
\affiliation{Time and Frequency Division, National Institute of Standards and Technology, 325 Broadway, Boulder, CO 80305, USA} 
\affiliation{Department of Physics, University of Colorado, 2000 Colorado Avenue, Boulder, CO 80309, USA}

\author[0000-0001-8127-5775]{Arpita Roy}
\affil{Space Telescope Science Institute, 3700 San Martin Dr., Baltimore, MD 21218, USA}

\author[0000-0002-4046-987X]{Christian Schwab}
\affil{Department of Physics and Astronomy, Macquarie University, Balaclava Road, North Ryde, NSW 2109, Australia}




\begin{abstract}
The coherent low-frequency radio emission detected by LOFAR from Gliese 1151, a quiescent M4.5 dwarf star, has radio emission properties consistent with theoretical expectations of star-planet interactions for an Earth-sized planet on a 1-5 day orbit. New near-infrared radial velocities from the Habitable-zone Planet Finder (HPF) spectrometer on the 10\,m Hobby-Eberly Telescope at McDonald Observatory, combined with previous velocities from HARPS-N, reveal a periodic Doppler signature consistent with an $m\sin i = 2.5 \pm 0.5 M_\oplus$ exoplanet on a 2.02-day orbit. Precise photometry from the Transiting Exoplanet Survey Satellite (TESS) shows no flares or activity signature, consistent with a quiescent M dwarf. While no planetary transit is detected in the TESS data, a weak photometric modulation is detectable in the photometry at a $\sim2$ day period. This independent detection of a candidate planet signal with the Doppler radial-velocity technique adds further weight to the claim of the first detection of star-exoplanet interactions at radio wavelengths, and helps validate this emerging technique for the detection of exoplanets. 
\end{abstract}



\section{Introduction} 
\label{sec:intro}
Planets in close-in orbits orbit are embedded in a magnetized stellar wind from the expanding stellar corona. As they orbit, short-period planets can perturb the flow of the magnetized wind, which can carry substantial amounts of energy towards the host star via sub-Alfvénic interactions \citep{neubauer1980,saur2013,turnpenney2018}. This incoming energy can heat up the chromosphere of the star, causing a hot spot on the surface of the star, which can cause variability that is modulated by the orbital period of the planet. \cite{shkolnik2005,shkolnik2008} and \cite{cauley2019} detected chromospheric modulations in hot Jupiter systems, which they interpret as evidence for Star-Planet Interactions (SPI), and \cite{turner2020} discuss promising candidate detections of circularly polarized emission from $\tau$ Boötis and $\nu$ Andromedae. In addition to hot Jupiters, M-dwarf planet systems with close-in rocky planets---such as the TRAPPIST-1 system \citep{gillon2017}---have also been suggested as capable of producing SPI \citep{pineda2018,turnpenney2018,fischer2019}. Recently, \cite{perez2021} announced the detection of circularly polarized 1.6 GHz radio emission from Proxima Centauri that could be consistent with sub-Alfvénic interactions with its 11.2-day planet. Continued monitoring at radio wavelengths will help to distinguish between sub-Alfvénic interactions and other mechanisms \citep[eg.][]{zic2020}.

\cite{vedantham2020} reported evidence for low-frequency highly circularly polarized radio emission at 150 MHz in GJ 1151 using the LOFAR Telescope Array \citep[the LOw-Frequency ARray;][]{haarlem2013}. GJ 1151 is a nearby bright ($J=8.5$) M4.5 dwarf with a quiet chromosphere. The origin of the radio emission, which is observed to be transient and highly circularly polarized (circular polarization fraction of $64 \pm 6\%$), is most consistent with sub-Alfvénic star-planet interactions between a rocky planet in a 1 to 5 day orbit around the host star.

To search for the presence of a potential rocky planet companion and exclude more massive companions, \cite{pope2020} obtained precise radial velocities (RVs) using the HARPS-N spectrograph. Their observations, yielding 19 RVs over a span of three months, allowed them to place upper limits on the mass of a potential planetary companion of $m\sin i < 5.6 \unit{M_\oplus}$ with periods between 1 and 5 days.

We report on a rocky planet candidate revealed in additional precise RVs obtained in the near-infrared using the Habitable-zone Planet Finder \citep{mahadevan2012,mahadevan2014} on the 10m Hobby-Eberly Telescope. Together, the HPF RVs and the HARPS-N RVs from \cite{pope2020} reveal a planet with an orbital period of $P=2.02 \unit{days}$ and an RV semi-amplitude of $K = 4.1 \pm 0.8 \unit{m\,s^{-1}}$, translating to a mass of $m\sin i = 2.5 \pm 0.5 M_\oplus$, where $i$ is the inclination of the orbit. Given its short period, the planet is capable of sub-Alfvénic interactions with its host star, and is likely source of the coherent radio emission observed by \cite{vedantham2020}. Photometric data from the Transiting Exoplanet Survey Satellite \citep[TESS;][]{ricker2015} confidently rules out transits of the planet candidate, but the data show hints of modulation at $\sim$2 days at the $100 \unit{ppm}$ level, which could constitute the photometric signature of the SPI interaction.

This paper is structured as follows. Section \ref{sec:data} details the observations. Section \ref{sec:analysis} presents the analysis of the HPF and HARPS-N RVs along with the TESS photometric data. In Section \ref{sec:discussion}, using the SPI model from \cite{saur2013} and \cite{turnpenney2018}, we demonstrate that the planet candidate satisfies the sub-Alfvénic criterion and is capable of SPI, and we further discuss the energetics of the sub-Alfvénic interactions. We summarize our key findings in Section \ref{sec:summary}.

\section{Data} 
\label{sec:data}

\begin{figure*}
\includegraphics[width=\textwidth]{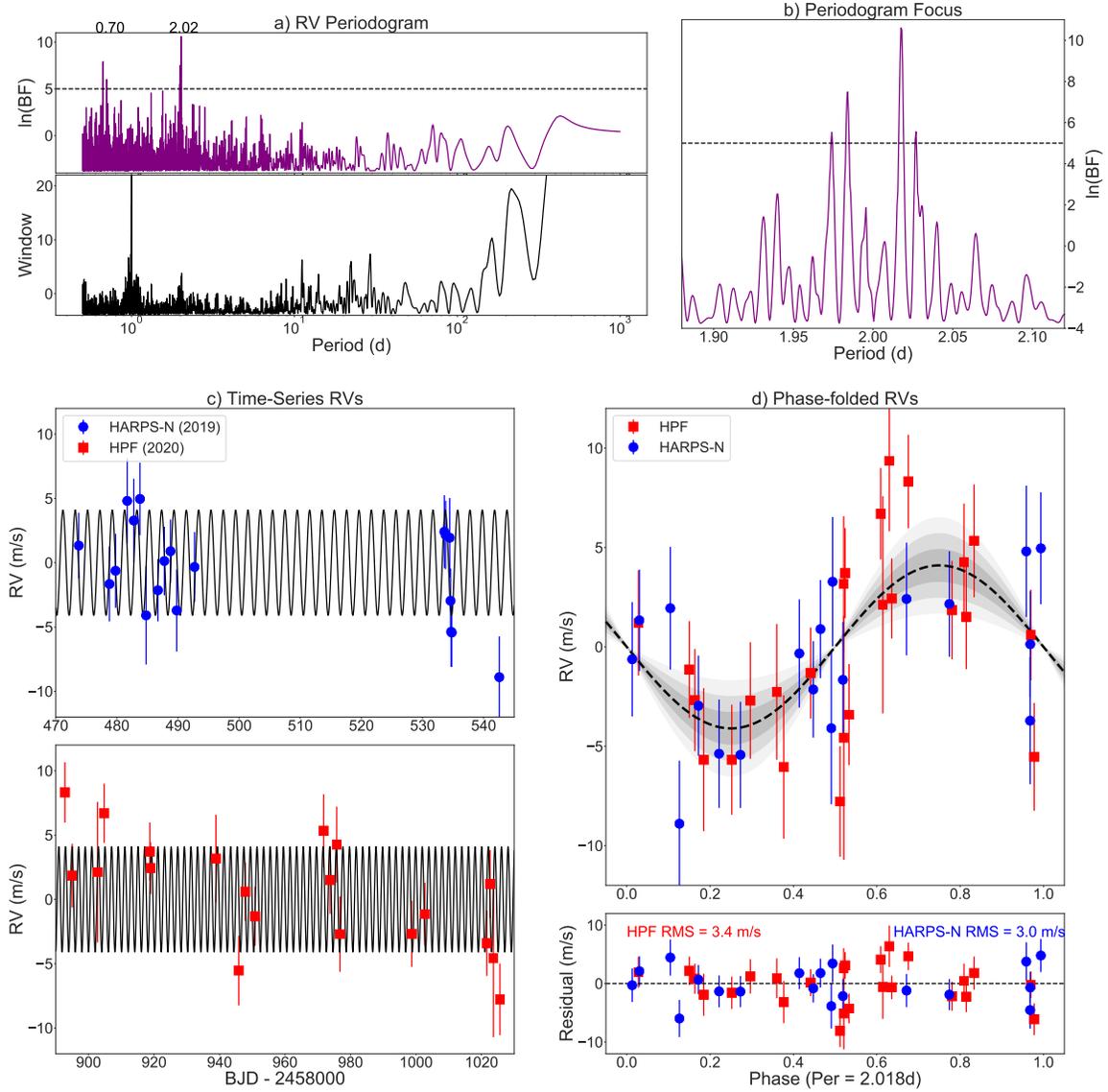}
\caption{HARPS-N and HPF radial velocities of GJ 1151. In \emph{a}, we show the Bayes factor periodogram of the combined HARPS-N+HPF RV series, along with the corresponding window function.  The section of the periodogram near the 2.02-day peak is detailed in \emph{b}.  In \emph{c}, we show the time-series RVs with the Keplerian model to the planet overlaid.  The 5 earliest HPF RVs are not shown to facilitate visibility. In \emph{d}, we show all RVs phased to the 2.018-day period of the planet. The planet model is shown as a black curve, with 1, 2, and 3-$\sigma$ uncertainty regions shaded in gray. RVs and associated activity indicators are available in machine readable format.}
\label{fig:rv}
\end{figure*}

\subsection{HARPS-N Radial Velocities}
\citet{pope2020} obtained time-series Doppler spectroscopy of GJ 1151 using the HARPS-N spectrometer \citep{cosentino2012} at Telescopio Nazionale Galileo (TNG). The HARPS-N time series consists of 21 observations compiled over a time baseline of approximately 3 months between December 2018 and February 2019.

The standard HARPS-N RV pipeline did not offer sufficient Doppler precision for a star as cool as GJ 1151, so \citet{pope2020} extracted their own RVs using the \wobble \citep{bedell2019} spectral analysis code.  From the 21 HARPS-N spectra, \citet{pope2020} provided 19 \wobble RVs; one spectrum (BJD=2458475.75494602) was explicitly excluded for having low signal-to-noise, while another (BJD=2458491.64178017) was excluded for unspecified reasons. The \wobble RVs exhibit an RMS scatter of 3.8 \mpsp, with a mean single-measurement error of 2.9 \mpsp.

In our analysis, we have incorporated the \wobble RVs as presented by \citet{pope2020}.  For our analysis of stellar magnetic activity (\S\ref{sec:activity}), we have also used the 1D extracted HARPS-N spectra from the TNG archive to derive the Ca II H\&K \citep[\shkp;][]{vaughan1978,gds2011} and \ha \citep[\ihap;][]{robertson2013} activity indices. 

\subsection{HPF Radial Velocities}
HPF is a stabilized fiber-fed near-infrared (NIR) spectrograph on the 10\,m Hobby-Eberly Telescope (HET) covering the $z$, $Y$, and $J$ bands from $810-1280 \unit{nm}$ at a spectral resolution of $R \sim 55000$. To enable precise RVs in the NIR, HPF is actively temperature stabilized to the milli-Kelvin level \citep{stefansson2016}. HET is a fully queue-scheduled telescope \citep{shetrone2007}, and all of the observations presented here were executed as part of the HET queue.

HPF uses a Laser-Frequency Comb (LFC) calibrator \citep{metcalf2019} to provide long-term accurate and precise instrument drift correction and monitoring. We did not obtain simultaneous LFC observations to minimize the possibility of contaminating the target spectrum, but performed the drift correction using LFC frames obtained throughout the night. We have previously demonstrated that we can achieve a drift-corrected precision at the $<30\unit{cm}\unit{s}^{-1}$ level \citep{stefansson2020} using this technique.

We used the \texttt{HxRGproc} \citep{ninan2018} package to perform bias, non-linearity, and cosmic-ray corrections, along with slope and variance image generation of the raw HPF up-the-ramp data. Following the slope and variance image generation, the HPF 1D spectra were extracted using the methods and algorithms discussed by \cite{ninan2018}, \cite{kaplan2019}, and \cite{metcalf2019}. To extract RVs from the 1D HPF spectra, we used an adapted version of the SpEctrum Radial Velocity AnaLyzer  \citep[SERVAL;][]{zechmeister2018} code which is further described by \cite{metcalf2019} and \cite{stefansson2020}. Briefly, SERVAL uses the template-matching method \citep{anglada2012} to extract precise RVs. For the RV extraction, we followed \cite{stefansson2020}, using the 8 HPF orders cleanest of tellurics covering the wavelength range from 854-889nm and 994-1076nm. Following \cite{metcalf2019} and \cite{stefansson2020}, we masked tellurics and sky-emission lines, to minimize their effect on the RV determination. To further minimize the impact of sky-emission lines on the RV determination, we subtracted the sky-background estimated using the dedicated HPF sky-fiber. Barycentric correction of our RVs was performed using \texttt{barycorrpy} \citep{kanodia2018}.  \texttt{barycorrpy} also corrects for secular acceleration \citep{kurster2003}, which for GJ 1151 is 0.61 $\unit{m}\unit{s}^{-1}\unit{yr}^{-1}$.  The secular acceleration is negligible across the baseline of the HARPS-N RVs, but significant for our HPF velocities.

Overall, we obtained 50 high resolution spectra with HPF in 25 visits with HPF, obtaining two spectra per visit with an exposure time of 969s per exposure. The spectra cover a time baseline of 458 days from March 15 2019 to June 25 2020. The median S/N of the HPF spectra was 217 per 1D extracted pixel at $1\unit{\mu m}$. We excluded from our analysis two of the spectra due to low S/N of 28 and 38 due to poor weather, which left 48 high quality spectra obtained in 24 visits. Following \cite{stefansson2020}, we binned the resulting RV points per HET visit, resulting in a median RV error of 2.7 \mpsp, and an effective exposure time of 32 minutes. The HPF RVs are shown alongside the HARPS-N velocities in Figure \ref{fig:rv}. The RVs and associated activity indicators used in this work are provided in machine readable format as `Data Behind Figure' \ref{fig:rv}.

\subsection{TESS Photometry}
As part of its all-sky survey for transiting exoplanets, the Transiting Exoplanet Survey Satellite \citep[TESS;][]{ricker2015} observed GJ 1151 for 27 days during Sector 22 (18 February--18 March 2020) of the mission in two orbits (Orbit 51 and 52). GJ 1151 is listed as TIC 11893637 in the TESS Input Catalog \citep[TIC;][]{stassun2018,stassun2019}. TESS pixel data surrounding GJ 1151 were averaged into 2-minute stacks, which were reduced to lightcurves by the Science Processing Operations Center (SPOC) at NASA Ames \citep{jenkins2016}, which we retrieved using the \texttt{lightkurve} package \citep{lightkurve}. We analyzed the Presearch Data Conditioning Single Aperture Photometry (PDCSAP) lightcurve, which contains systematics-corrected data using the algorithms originally developed for the \textit{Kepler} data analysis pipeline. The PDCSAP lightcurve uses pixels chosen to maximize the SNR of the target and has removed systematic variability by fitting out trends common to many stars \citep{smith2012,stumpe2014}. Data from \textit{Gaia} demonstrate that there are no nearby stars within 1 arcmin that are within 5 TESS magnitudes of GJ 1151, resulting in minimal dilution of the TESS light curve.

To clean the available TESS data, we removed all points with non-zero quality flags (4563 in total) which indicate known problems \citep[e.g.,][]{tenenbaum2018}. We removed an additional 6 points that we identified as $4\sigma$ outliers, leaving a total of 14050 points that we used for subsequent analysis, with a median errorbar of 1392ppm. The median-normalized TESS $\mathrm{PDCSAP}$ light curve is discussed in Subsection \ref{sec:tess}. 

From the TESS Data Release Notes for Sector 22, 'momentum dump' events---when the TESS reaction wheel speeds are reset by removing angular momentum through thruster firing to keep the pointing of the telescope stable---occurred every 6.625 days, and 6.75 days in Orbits 51, and 52, respectively. These events are known to impact the photometry \citep[see e.g.,][]{huang2018}, and we specifically highlight those events in Figure \ref{fig:tess}.

\section{Analysis} 
\label{sec:analysis}

\subsection{Period Search in RV Data}

When analyzing the combined HARPS-N+HPF RVs with frequency analysis tools such as periodograms, we consistently find evidence for a periodic signal near 2 days.  For the sake of brevity, we will present results from the Bayes Factor Periodogram, or BFP \citep{feng2017}.  The BFP is particularly useful for this analysis, as it offers an unambiguous measure of a signal's statistical significance, which is crucial when attempting to detect a low-amplitude exoplanet with a small number of observations.  However, other periodograms such as the generalized Lomb-Scargle \citep[GLS;][]{zk09} or Bayesian Lomb-Scargle \citep[BGLS;][]{mortier2015} offer qualitatively similar results.

The BFP works by comparing the Bayesian Information Criterion (BIC) of a periodic signal at each candidate frequency with that of a noise model.  Periodogram peaks with a Bayes factor $\ln(BF) \geq 5$ are considered statistically significant.  While the \texttt{Agatha} implementation of this algorithm from \citet{feng2017} offers the option to account for correlated noise using one or more moving average terms, the model selection feature of \texttt{Agatha} prefers a white noise model for our data.  Our BFPs were all evaluated for periods $-0.3 < \log_{10}(P) < 3$ with an oversampling factor of 15.

In Figure \ref{fig:rv}a, we show the BFP for the combined HARPS-N+HPF time series.  The RV periodogram shows two distinct sets of peaks.  The first occurs at around 2 days, with peaks at 2.02 and 1.98 days.  The second feature sits at approximately 0.7 days.  These peaks are all aliases of each other; specifically, each frequency is separated by 1 day$^{-1}$. The ``1-day alias" problem is common to time-series Doppler searches \citep{dawson2010}.  Interestingly, the ambiguity between 1.98 and 2.02 days is extremely similar to the alias observed for YZ Ceti b, another low-mass exoplanet orbiting a mid-M dwarf \citep{robertson2018}.  The periodograms suggest that 2.02 days is the true signal period, a hypothesis we confirm via model comparison in \S\ref{sec:modeling}.

The HPF RVs alone show statistically significant power near 2 days.  The HARPS-N velocities alone do not show significant power at any period, but the strongest peak occurs near the 0.7-day alias of the 2-day period.  While the power in the combined periodogram is approximately equal to that of the HPF power spectrum, the appearance of at least one alias of the same signal in HARPS-N, as well as the consistency of the HARPS-N RVs with the 2-day signal (Figure \ref{fig:rv}c) would appear to rule out instrumental systematics as the origin of the signal.

\subsection{Ruling out an Astrophysical False Positive}
\label{sec:activity}
Magnetic features such as starspots and plage can create periodic RV signals \citep[e.g.,][]{boisse2011,robertson2014}, and the 2-day period of our candidate signal would be consistent with many young, rapidly rotating M dwarfs \citep{newton2016}.  However, we find instead that GJ 1151 appears to be quiet and slowly-rotating, and that its rotation should not be the astrophysical origin of the signal.

By all indications, GJ 1151 is a slow rotator. \citet{newton2016} estimated its rotation period to be 117.6 days based on MEarth photometry.  \citet{reiners2018} placed an upper limit on its rotational velocity of $<2$ km\,s$^{-1}$, which we agree with based on our own $v\sin i$ analysis of the high-resolution HPF spectra, from which we also obtain an upper limit of $v\sin i < 2 \unit{km s^{-1}}$. Its X-ray luminosity $L_X = 5.5 \times 10^{26}$ erg\,s$^{-1}$ \citep{foster2020} implies a stellar age of approximately 5 Gyr and a rotation period between 70 and 90 days, according to the empirical relationship of \citet{engle2011}. Furthermore, data from TESS, HARPS-N, and HPF disfavor a short rotation period. For the HARPS-N and HPF time series, we performed periodicity searches for four spectral activity indicators: \shk and \iha from HARPS-N, and the chromatic index (CRX) and differential line width (dLW) from HPF \citep{zechmeister2018}.  We see no evidence for periodic astrophysical variability in any of the spectral activity tracers.  Likewise, the TESS lightcurve shows no evidence for a stellar rotation period of $P \leq 15$ days. Finally, the complete absence of detectable flares in the TESS lightcurve, and the lack of emission in chromospheric lines such as \hap, are inconsistent with GJ 1151 being young and active. Further still, the kinematics of this star are consistent with an older age as would be expected for a slow rotator: $UVW$ = (-26.9, -65.0, -33.4)~km~s$^{-1}$, placing it at the boundary of thin and thick disk membership.

Aside from arguments related to the difference between the star's likely rotation period and the RV period, the RV data themselves are inconsistent with activity-driven variability. The RVs show no correlations with any of the spectral activity indicators. Also, if we model the 2.02-day signal using the optical HARPS-N RVs and NIR HPF RVs separately, we find consistent amplitudes. On the other hand, if the signal were created by starspots, we would expect a smaller amplitude at NIR wavelengths \citep[e.g.,][]{marchwinski2015}. Thus, we conclude that it is extremely unlikely that the observed 2-day Doppler signal is caused by stellar variability.

\begin{table*}[t!]
\begin{center}
\begin{tabular}{| l c  c |}
\hline
Parameter & Prior & Posterior \\
\hline
& & \\
\multicolumn{3}{| c |}{\emph{Orbital Parameters}} \\
                                                     &                                     &                                 \\
Period $P$ (days)                                    &  $\mathscr{N}(2.016, 0.1)$          & $2.0180 \pm 0.0005$             \\
Time of inferior conjunction $T_C$ (BJD - 2458400)   &  $\mathscr{U}(72.6655, 74.6835)$    & $73.691 \pm 0.07$               \\
$\sqrt{e}\sin\omega$                                 &  $\cdots$                           & 0 (fixed)                       \\
$\sqrt{e}\cos\omega$                                 &  $\cdots$                           & 0 (fixed)                       \\
RV semi-amplitude $K$ (\mpsp)                        &  $\mathscr{U}(0, 100)$              & $4.1 \pm 0.8$                   \\
                                                     &                                     &                                 \\
                                                     &                                     &                                 \\
\multicolumn{3}{| c |}{\emph{Instrument Parameters}} \\
                                                     &                                     &                                 \\
HARPS-N zero-point offset $\gamma_{\mathrm{HARPSN}}$ (\mpsp)  &  $\mathscr{U}(-20, 20)$             & $0.7 \pm 0.7$                   \\
HPF zero-point offset $\gamma_{\mathrm{HPF}}$ (\mpsp)         &  $\mathscr{U}(-20, 20)$             & $-0.4 \pm 0.7$                  \\
HARPS-N jitter $\sigma_{\mathrm{HARPSN}}$ (\mpsp)             &  $\propto (\sigma + \sigma_0)^{-1}$ & $0.4^{+1.0}_{-0.4}$             \\
HPF jitter $\sigma_{\mathrm{HPF}}$ (\mpsp)                    &  $\propto (\sigma + \sigma_0)^{-1}$ & $1.8 \pm 1$                     \\
                                                     &                                     &                                 \\
                                                     &                                     &                                 \\
\multicolumn{3}{| c |}{\emph{Inferred Parameters}}   \\
                                                     &                                     &                                 \\
Minimum mass $m \sin i$ ($M_\oplus$)                   &                                     & $2.5 \pm 0.5$                   \\
Semimajor axis $a$ (AU)                              &                                     & $0.01735_{-0.00070}^{+0.00065}$ \\
Semimajor axis $a$ ($a/R_*$)                         &                                     & $18.5 \pm 0.9$                  \\
                                                     &                                     &                                 \\
\hline
\end{tabular}
\caption{Priors and 1-dimensional posterior distributions for the orbital and instrumental parameters for our 1-planet RV model to GJ 1151. $\mathscr{N}(a,b)$ denotes a normal prior with median value $a$, and standard deviation $b$; $\mathscr{U}(a,b)$ denotes a uniform prior with lower limit value $a$ and upper limit value $b$.}
\label{tab:orbit_params}
\end{center}
\end{table*}

\subsection{Orbit Modeling}
\label{sec:modeling}
Given the evidence that GJ 1151 is a quiet, slowly-rotating star, we modeled the 2-day variability as a Keplerian exoplanet orbit. We used the Markov Chain Monte Carlo (MCMC) orbit-fitting code \radvel \citep{fulton2018} to compute the model. We adopted mostly uninformative priors for the model parameters, although for the additional white-noise terms (``jitters", $\sigma_{\mathrm{HARPSN/HPF}}$) we adopted the jitter prior suggested by \citet{ford2007} with a ``reference value" $\sigma_0$ = 1 \mps and maximum $\sigma_{\textrm{max}} = 100$ \mpsp. This prior was chosen to facilitate Bayesian model comparison, particularly to a zero-planet model, for which a uniform prior can allow the noise terms to grow enough to absorb real astrophysical variability. We fixed the planet's eccentricity to zero, both because our data are not numerous enough to constrain it, and because we expect a planet so close to the star to be tidally circularized \citep[e.g][]{rasio1996}.

For each model considered, we compared to a zero-planet model in which we account for the RV variability using only a zero-point offset and a jitter term for each instrument. We performed model selection using the Bayes factor $\ln{BF} = \frac{\Delta BIC}{2}$, where $BIC$ is the Bayesian Information Criterion, which scales a model's log-likelihood value to include a complexity penalty for models with more free parameters \citep{kass1995}.

Our preferred model is for an exoplanet with $P = 2.02$ days. When compared to the zero-planet model, it has a Bayes factor $\ln{BF} = 12.7$, indicating a high level of significance. This Bayes factor also agrees well with the result of our periodogram analysis.  While the alias of the 2.02-day period at $P = 1.98$ days appears as a strong peak in our periodogram analysis, if we choose a period prior tightly constrained at 1.98 days, the posterior distribution still prefers the 2.02-day period. Additionally, we also note that while a model treating the signal as a Keplerian with $P = 0.7$ days is also favored over the zero-planet model, its Bayes factor is significantly lower than that of the 2-day model, and we therefore conclude it is simply an alias of the true period.

Given the multi-modal nature of the period posterior, we additionally fit the RV datasets using the \texttt{dynesty} dynamic nested sampler \citep{speagle2019} available in the \texttt{juliet} \citep{Espinoza2018} package.  Nested samplers are efficient at accurately exploring multi-modal solutions \citep{speagle2019}. From the nested sampler, we observe a clear highest mode at 2.02 days ($P=2.0183_{-0.0008}^{+0.0084} \unit{days}$), with a significantly smaller mode seen at 1.98 days. To quantify the preference for the 2.02 day solution over the 1.98 day solution we ran two sets of 6 RV-fits each in \texttt{juliet} to get an accurate view of the resulting variance in log evidence values ($\ln Z$ values). To sample the 2.02 day solution, we ran the first set with a uniform prior on the period from 2.0 to 2.04 days, and to sample the 1.98 day solution, we ran the second set with a uniform prior from 1.96 days to 2.0 days on the period. In doing so, we obtain a $\ln Z = -136.22 \pm 0.37$ and $\ln Z = -139.27 \pm 0.22$ for the 2.02 day and 1.98 day solutions, respectively, where we report the values as the median value from the 6 independent fits and the error as the standard deviation from the 6 fits. We adopt the 2.02 day solution, given the $\Delta \ln Z = 3$ (20 to 1 posterior odds) statistical preference for that solution over the 1.98 day solution. 

The planet candidate presented here is fully consistent with the expected planet occurrence around mid M-dwarfs: \cite{hardegree2019} suggest a planet occurrence of $1.4_{-1.0}^{+2.3}$ for small ($0.5 \unit{R_\oplus}$ to $2.5 \unit{R_\oplus}$) short-period ($P<10\unit{days}$) planets around M4-M4.5 dwarfs. There remains a possibility that further planets orbit in the system. However, as seen from Figure \ref{fig:rv} with the current RVs, we do not see clear evidence of additional periodic signals. Additional precise RVs could shed further constraints on any additional planets in the system.

The prior distributions and adopted posterior values of the planet's orbital model are shown in Table \ref{tab:orbit_params}.  Additionally, we show the orbit superimposed over the RV data in Figure \ref{fig:rv}.

\begin{figure*}[t!]
\centering
\includegraphics[width=1.0\textwidth]{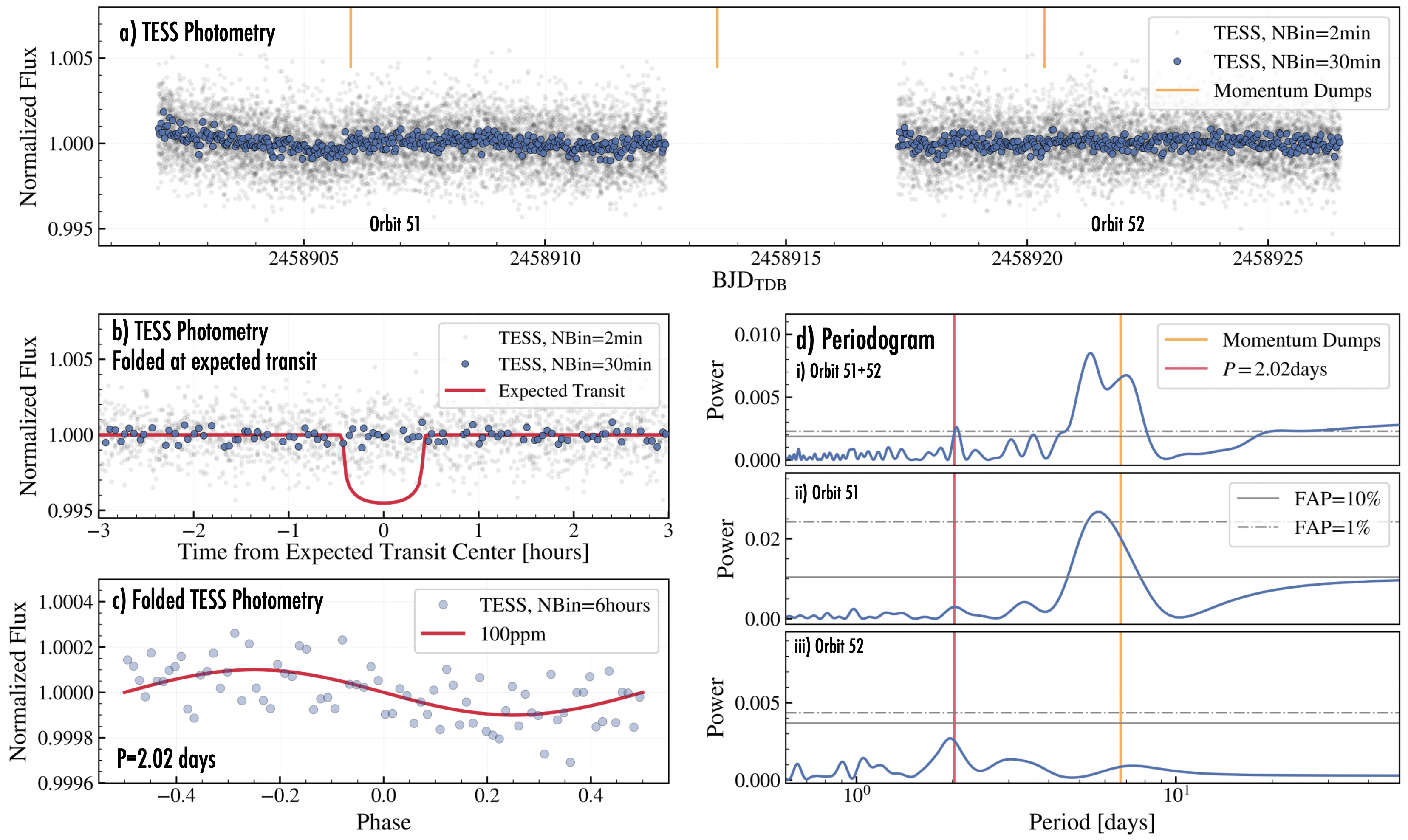}
\vspace{0.1cm}
\caption{a) Available photometry from TESS from Sector 22, obtained in two orbits: TESS Orbit 51 and 52. The unbinned 2min TESS cadence is shown in black, and 30min bins in blue. No transits are seen. b) TESS photometry folded on the expected transit ephemeris from our best-fit RV fit. The red curve shows a nominal expected transit model if the planet were transiting. The data rule out such transits to a high degree of confidence. c) TESS photometry of both orbits phased to the 2.02day period of the RV planet shows a $\sim$100ppm sinusoidal amplitude. d) Lomb-Scargle Periodograms of i) all available TESS photometry, ii) photometry from Orbit 51, and iii) photometry from Orbit 52. The data show clear peaks in and around the periodicity of the momentum dumps (6-7days), and show hints of a moderately significant peak (False Alarm Probability $\sim$1\%) at the 2.02day RV planet period.}
\label{fig:tess}
\end{figure*}

\subsection{TESS Photometry}
\label{sec:tess}

\subsubsection{Transit Search}
Given the short orbital period period of the planet, and the small radius of the host star, the planet has a high geometric transit probability. From our best-fit RV model, assuming a circular orbit, we estimate a semi-major axis of $a=0.017 \unit{AU}$. Using a stellar radius of $R=0.2016 R_\oplus$, we obtain a geometric transit probability of $R_*/a = 5.4\%$. Thus, we looked for evidence of transits at the expected times inferred from our best-fit RV solution in photometry available from TESS.

Figure \ref{fig:tess}a shows the available photometry from TESS, and Figure \ref{fig:tess}b shows the expected transit of GJ 1151b phased at the best-fit ephemerides from our RV fit. To estimate the expected transit depth, we predicted the most likely radius of the planet from its minimum mass value of $m_{\mathrm{min}} = 2.5 M_\oplus$ (assuming inclination is $i=90^\circ$) using the parametric mass-radius relation implemented in the \texttt{forecaster} \citep{chen2017} package. Using \texttt{forecaster}, we obtain a radius of $R=1.35_{-0.29}^{+0.53} R_\oplus$, which corresponds to an expected transit depth of $\sim3800 \unit{ppm}$. As seen from Figure \ref{fig:tess}b, the expected transit depth of the planet is significantly deeper than TESS's single-point 2min photometric precision on this target ($1200 \unit{ppm}$), and the data confidently rule out transits at this ephemeris. In addition, to search for evidence of other periodic transiting planets in the system, we used the Box-Least-Squares (BLS) algorithm \citep{kovacs2002}. In examining BLS periodograms from periods from 0.5 to 30 days, we see no significant evidence of transiting planets in the system from the TESS data. We additionally note that although sub-Alfvénic star-planet interactions are capable of creating flares, we see no clear signatures of flares in the TESS data \citep{fischer2019}.

\subsubsection{Periodogram Analysis}
Figure \ref{fig:tess}d shows Generalized Lomb Scargle (GLS) Periodograms of the available TESS photometry, showing the GLS of i) all of the available photometry (Orbits 51 and 52; top panel), ii) photometry from Orbit 51 (middle panel), and iii) photometry from Orbit 52 (bottom panel). In Orbit 51, we see a clear peak at $\sim6-7 \unit{days}$, which we attribute to the systematic noise in the photometry that occurred during the TESS momentum dump in this orbit (see Figure \ref{fig:tess}a). From the GLS Periodogram of Orbit 52, which overall shows less correlated noise structures, we do not see a clear systematic jump during the TESS momentum dump, and we see no clear peak at 6-7 days. In the GLS Periodogram of Orbit 52, and in the periodogram of both orbits, we see evidence of a moderately significant peak at the $2.02 \unit{day}$ period we see in the RVs (False Alarm Probability of $\sim$1\%), which corresponds to a $\sim$100ppm photometric signal after phasing all of the available photometry to the 2.02day period (Figure \ref{fig:tess}c).

To investigate the possibility that this 100ppm photometric signal is due to ellipsoidal variations from the orbiting planet, we estimated the expected amplitude of ellipsoidal variations using Equation 7 of \cite{shporer2017}. Using our $a/R_* = 18.5$ and assuming $m_p \sin i = 2.5 M_\oplus$, we obtain a maximum possible amplitude expected from ellipsoidal variations of $\sim$$0.01\unit{ppm}$, which is negligible in comparison to the observed $100 \unit{ppm}$ signal in TESS. We conclude that the $100\unit{ppm}$ photometric modulation is not due to ellipsoidal variations. 

Instead, we surmise that the 2 day photometric signal could represent the photometric counterpart of the star-planet interaction between the planet and the star. However, due to the limited significance of the signal in the TESS photometry (False Alarm Probability of $\sim$1\% in Figure \ref{fig:tess}), we urge further precise photometric follow-up of this system to  characterize this potential low-amplitude signal.

\begin{table*}[t!]
\centering
\caption{Summary of input parameters and posterior parameters describing the energetics of the star-planet interactions for the ST model (\citealt{saur2013}, and \citealt{turnpenney2018}), and the Lanza model \citep{lanza2009}. $\mathscr{N}(a,b)$ denotes a normal prior with median value $a$, and standard deviation $b$; $\mathscr{J}(a,b)$ denotes a log-uniform prior with lower limit value $a$ and upper limit value $b$; $\mathscr{U}(a,b)$ denotes a uniform prior with lower limit value $a$ and upper limit value $b$.}
\begin{tabular}{llcl}
\hline\hline
Parameter                           & Parameter Description                       & Value                            & Notes  \\
\hline
\multicolumn{4}{l}{~~~\emph{Prior Parameters}} \\
$R_*$                                 & Stellar Radius ($R_\odot$)                & $\mathscr{N}(0.2016, 0.0060)$    & TICv8 \citep{stassun2019} \\
$M_*$                                 & Stellar Mass ($M_\odot$)                  & $\mathscr{N}(0.171, 0.020)$      & TICv8 \citep{stassun2019} \\
$P_{\mathrm{rot}}$                    & Stellar Rotation (days)                   & $117$                            & \cite{newton2016} \\
$T_{\mathrm{C}}$                      & Temperature of Stellar Corona (K)         & $2 \times 10^{6}$                & Adopted from \cite{vedantham2020} \\
$n_{\mathrm{base}}$                   & Base Number Density ($\mathrm{cm^{-3}}$)  & $1\times 10^6$                   & Adopted from \cite{vedantham2020} \\
$B_*$                                 & Stellar Magnetic Field (T)                & $\mathscr{J}(0.001, 0.1)$        & Nominal M-dwarf stellar magnetic field \\
$B_{\mathrm{exo}}$                    & Planet Magnetic Field (G)                 & $\mathscr{J}(0.1,10)$            & Nominal planet magnetic field \\
$R_{\mathrm{p}}$                      & Planet Radius ($R_\oplus$)                & $\mathscr{N}(1.5, 0.5)$          & Estimated from $m\sin i$ \\
\multicolumn{4}{l}{~~~\emph{Posterior Parameters}} \\
$\log_{10}(M_{\mathrm{A}})$           & Alfvén Mach number                        & $-1.6 \pm 0.7$                   &   \\
$\log_{10}(S_{\mathrm{total,ST}})$    & Poynting Flux, ST model ($\mathrm{ergs\,s^{-1}}$)          & $22.0_{-0.6}^{+0.7}$             &   \\
$\log_{10}(S_{\mathrm{total,Lanza}})$ & Poynting Flux, ST model ($\mathrm{ergs\,s^{-1}}$)          & $24.0_{-0.9}^{+1.1}$             &   \\
\hline
\end{tabular}
\label{tab:Alfvénmodel}
\end{table*}

\subsubsection{Starspot Modulation}
Given that we have identified photometric periodicity matching that of our proposed exoplanet candidate, we considered the possibility that the RV signal is caused by starspot modulation rather than a planet. This spot could be created by H- opacity supplied by the incoming electron beam from the possible sub-Alfvénic interaction, similar to stellar spots observed in photometry of late M-dwarfs with corresponding radio modulations \citep[see e.g.,][]{littlefair2008,hallinan2015}. To estimate the spot size from the 100ppm TESS signal, we created a starspot model with \texttt{SOAP 2.0} \citep{dumusque2014}, which predicts both photometric and RV variability for given stellar parameters and starspot configurations. For the purposes of this test, we assumed the stellar rotation period $P_{\mathrm{rot}} = 2.02$ days. We used a simple model of a single equatorial starspot with a spot-photosphere temperature difference $\Delta T = 600$ K, which is a moderate contrast for M stars \citep{reiners2010}. Scaling the spot radius to match the observed photometric amplitude, we found that a radius equal to 1\% of the stellar radius produces the desired 100 ppm brightness variation. However, the expected RV amplitude of such a spot is of order 0.2 \mpsp, which is too small by a factor of $\sim$20 to explain the RV periodicity we observe. Furthermore, as we have discussed in \S\ref{sec:activity}, multiple lines of evidence suggest that $P_{\mathrm{rot}}$ is significantly longer than 2 days. Thus, we find it is unlikely that starspot modulation is the origin of either the photometric or Doppler signal at 2 days. 

\section{Discussion} 
\label{sec:discussion}

\subsection{Sub-Alfvénic Interaction}
To study the Alfvén interaction of the planet candidate and estimate the resulting Poynting fluxes, we broadly follow \cite{vedantham2020}, using the model frameworks of \cite{saur2013} and \cite{turnpenney2018}, and that of \cite{lanza2009}, to describe the energetics of the star-planet interactions. These models assume that the planet is a conductive perturber orbiting in a magnetized expanding stellar corona described as a Parker-wind \citep{parker1958}. For the Parker-wind, we assumed a coronal temperature of $T_{\mathrm{corona}} = 2\times 10^{6} \unit{K}$, which results in a sound speed of $c_{s} \sim 129 \unit{km}\unit{s}^{-1}$. Following \cite{vedantham2020}, we assumed a base number density of $n_{\mathrm{base}} = 10^6 \unit{g}\unit{cm}^{-3}$, which we scaled as $n \propto d^{-2}$ where $d$ is the distance from the star. 

To estimate if our planet candidate satisfies the criterion for sub-Alfvénic interactions, we estimate the Alfvén Mach number of the planet, which is given by,
\begin{equation}
M_A = v_{\mathrm{rel}}/v_A,
\end{equation}
where $v_{\mathrm{rel}}$ is the relative velocity of the stellar wind as seen by the orbiting planet, and $v_A$ is the Alfvén speed. Assuming a circular orbit with a period of $P=2.02 \unit{days}$, the planet has a Keplerian orbital velocity of $v_{\mathrm{orb}} \sim 94 \unit{km}\unit{s}^{-1}$ at its orbital distance of $a/R_* = 18.5$. As the magnetic field of GJ 1151 is not currently well constrained, if we adopt the nominal value of $B_{*} = 0.01 \unit{T}$ assumed by \cite{vedantham2020}, we obtain an Alfvén speed of $v_A \sim 11800 \unit{km}\unit{s}^{-1}$ and an Alfvén Mach number of $M_A = 0.026$ at the orbital distance of the planet. As the Mach number is less than 1, the planet is capable of sub-Alfvénic interactions with its host star. We further note the planet also satisfies the second criterion for sub-Alfvénic interactions -- that the radial wind speed is less than the radial component of the Alfvén speed, which is a necessary condition so that one of the two Alfvén-wings points towards the star \citep[see discussion in][]{saur2013}.

As the planet is capable of sub-Alfvénic interactions, and has a small Alfvén Mach number, we 
estimate the total Poynting flux of the system with
\begin{equation}
S_{\mathrm{total}} = 2 \pi \frac{v_{\mathrm{rel}} R_{\mathrm{eff}}^2 B_0^2}{\mu_0} \varepsilon,
\end{equation}
where $v_{\mathrm{rel}}$ is the relative velocity of the stellar wind as seen by the orbiting planet, $B_0$ is the total magnetic field at the position of the planet, $\mu_0$ is the vacuum permeability constant, and $R_{\mathrm{eff}}$ is the effective radius of the planet. 
The $\varepsilon$ parameter describes the efficiency of the star-planet interactions, and its parameterization differs in the model of \cite{saur2013} and \cite{turnpenney2018} (hereafter in the ST Model), and the model of \cite{lanza2009} (hereafter the Lanza Model). In the ST model $\varepsilon$ parameter is given by
\begin{equation}
\varepsilon = M_A \Bar{\alpha}^2 \sin^2(\theta),
\end{equation}
where $M_A$ is the Alfvén Mach number of the system, $\theta$ is the angle between the total magnetic field $B_0$ at the position of the planet and the total stellar wind velocity, and $\Bar{\alpha}$, is a parameter denoting the sub-Alfvénic interaction efficiency, where we follow \cite{vedantham2020} and assume it is equal to unity.

In the model of \cite{lanza2009}, $\varepsilon$ is given by
\begin{equation}
\varepsilon = \gamma /2,
\end{equation}
where $\gamma$ is a parameter ranging from 0 to 1. Here we follow \cite{vedantham2020} and assume $\gamma = 0.5$.

To estimate the effective radius of the planet, given our mass estimate of $m \sin i = 2.5 \pm 0.5 M_\oplus$, we use the \texttt{forecaster} \citep{chen2017} mass-radius relations to predict a likely minimum radius of the planet of $R_p = 1.35_{-0.29}^{+0.53} R_\oplus$. If the planet has an inherent magnetic field, this will increase the effective radius $R_{\mathrm{eff}}$ of the interaction \citep[see Equation 57 in][]{saur2013} by
\begin{equation}
R_{\mathrm{eff}} = R_{\mathrm{exo}} \left( \frac{B_{\mathrm{exo}}}{B_0} \right)^{1/3} \sqrt{3 \cos \left( \frac{\Theta_M}{2} \right) },
\end{equation}
where $R_{\mathrm{exo}}$ is the radius of the planet, $B_{\mathrm{exo}}$ is the equatorial magnetic field of the planet, and $\Theta_M$ is the angle between the planet magnetic moment and the stellar magnetic field at the location of the planet. We follow \cite{vedantham2020}, and assume that the planet has a magnetic field of $1 \unit{G}$, and assume that the magnetic moment nominally has a $\Theta_M = 90^\circ$ angle between the stellar magnetic field, resulting in $\sqrt{3 \cos ( \frac{\Theta_M}{2} )}\sim$1.46. As planets on short-period orbits are expected to be tidally locked to their stars, we make the assumption that the planet rotation period is equal to its orbital period, which we use to scale the resulting planetary magnetic field.

From these model input parameters, we obtain a Poynting flux of $S_{\mathrm{total,ST}} \sim 8 \times 10^{14} \unit{W}$, and $S_{\mathrm{total,Lanza}} \sim 9 \times 10^{16} \unit{W}$ for the ST and Lanza models, respectively. Assuming a Poynting flux-to-radio emission conversion efficiency of $\epsilon_r = 1\%$, we obtain a Poynting radio power of $P_{\mathrm{radio,ST}} = 8 \times 10^{12} \unit{W}$, and $P_{\mathrm{radio,Lanza}} = 9 \times 10^{14} \unit{W}$, for the two models, respectively. Given the distance of $d=8.04 \unit{pc}$ to the host star, we can also calculate the resulting spectral flux density of $F_{\mathrm{radio,ST}} = 0.5 \unit{mJy}$, and $F_{\mathrm{radio,Lanza}} = 52 \unit{mJy}$, where we followed followed \cite{vedantham2020} assuming a bandwidth of $\Delta \nu$ equal to the electron gyrofrequency on the surface of the star, and assumed a beam solid angle $\Omega = 0.1 \unit{sr}$. These spectral flux densities broadly agree with the observed LOFAR value of $\sim$$0.9\unit{mJy}$ from \cite{vedantham2020}. As a comparison to other M-dwarf systems, this spectral flux density is similar---but somewhat higher---to the spectral flux density of $\sim$0.3$\unit{mJy}$ for the M4-dwarf planet GJ 876 b as estimated by \cite{turnpenney2018}.

\begin{figure}[t!]
\centering
\includegraphics[width=0.95\columnwidth]{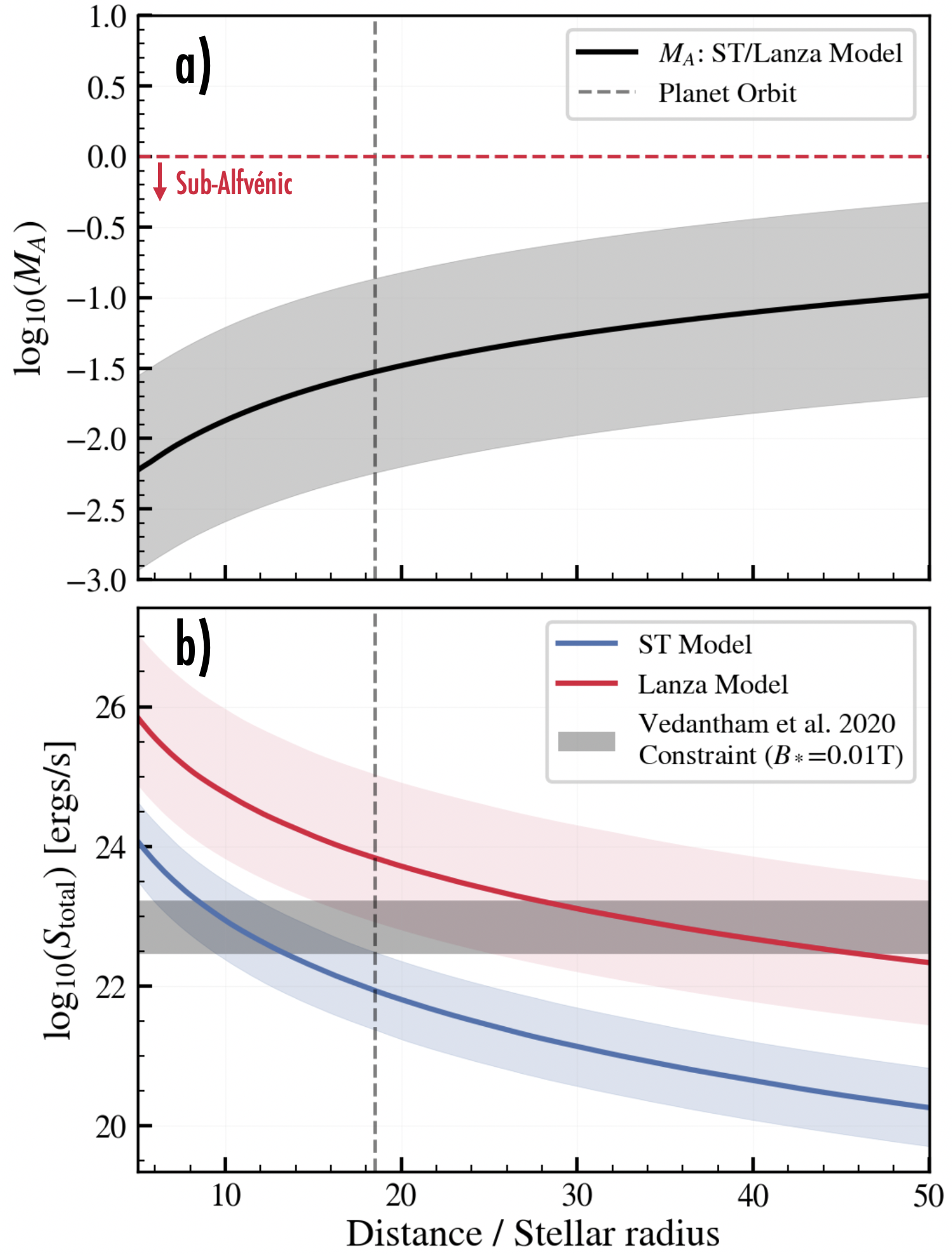}
\caption{a) Alfvén Mach numbers as a function of distance from the host star. The solid curve shows the median model, and the grey filled region shows the $1\sigma$ credible interval around the median model. The grey vertical dashed lines show the orbital distance of the planet. b) Total Poynting flux as a function of distance of the star as calculated with the ST model (blue solid curve) and the Lanza model (red solid curve). The blue and red shaded regions show the corresponding $1\sigma$ credible intervals around the median model. The grey region in the lower panel shows the radio constraint from \cite{vedantham2020} assuming a fixed $B_*=0.01 \unit{G}$.}
\label{fig:alfven}
\end{figure}

To visualize the sensitivity of the models to different input parameters as a function of orbital distance of the planet, we performed a Monte Carlo sampling assuming nominal input priors summarized in Table \ref{tab:Alfvénmodel}. Figure \ref{fig:alfven} shows the Alfvén mach number, along with the resulting Poynting flux estimates using the ST and the Lanza models, as a function of distance from the star. The solid curves show the median models and the corresponding shaded regions show the $1\sigma$ credible intervals. From Figure \ref{fig:alfven}a, we see that for even the broad set of input parameters in Table \ref{tab:Alfvénmodel}, the system is sub-Alfvénic ($\log_{10} (M_A) < 1$) at the orbital location of the planet. In Figure \ref{fig:alfven}b, we additionally compare our resulting Poynting flux estimates from the ST and Lanza models to the radio observation constraint from \cite{vedantham2020} (grey shaded region in Figure \ref{fig:alfven}b). We see that the ST and Lanza models flank the radio constraint presented by \cite{vedantham2020} which assumes a $B_* = 0.01 \unit{T}$, with both the ST and Lanza models being consistent with the observed radio constraint at the 1-2$\sigma$ level for the assumed parameters. We conclude that the planet candidate we report here is compatible with being the source of the radio emission detected by \cite{vedantham2020}.

In the near future, The Square Kilometre Array (SKA), a next generation radio telescope, is expected will come online which is anticipated to improve on the flux density sensitivity to that of LOFAR by a factor of 10-30 resulting in a flux density sensitivity of $\sim$10$\unit{\mu Jy}$ \cite{zarka2015} in the 50-250MHz range. With its improved precision, SKA is expected to enable the detection of additional planets exhibiting coherent radio emission around nearby stars.

\section{Summary}
\label{sec:summary}
We report on a planetary companion orbiting the quiescent M4.5 dwarf GJ 1151 in a 2.02 day orbit. From radial velocities obtained with the Habitable-zone Planet Finder and the HARPS-N spectrograph, we constrain the planet minimum mass to be $m\sin i = 2.5 \pm 0.5 M_\oplus$. The planet has an Alfvén Mach number of $M_A = 0.026$, and thus satisfies the criterion for sub-Alfvénic interaction with its host star. We estimate the resulting Poynting fluxes of the star-planet system using two different models, which we show are consistent with the radio constraints reported by \cite{vedantham2020} at the $1-2\sigma$ level. Given this consistency, we conclude that it is highly likely that the planet is the source of the radio emission. To confirm that SPI interactions with the RV planet candidate we report is the true source of the radio emission, we urge continued radio follow-up observations to demonstrate a corresponding 2.02day periodicity at radio wavelengths.

Further, using data from the TESS spacecraft we are able to rule out transits of the reported RV planet. From the TESS photometry, we see a photometric modulation at $\sim$2 days, with an amplitude of $\sim$100ppm, which could potentially be due to a spot on the surface of the star created by incoming electron beams supplied by the star-planet interactions.

This detection encourages further synergies of precision RVs and low-frequency radio observations with facilities such as LOFAR and SKA in the future to detect and characterize planets around nearby stars exhibiting coherent radio emission.

\acknowledgments
We thank the anonymous referee for a thoughtful reading of the manuscript, and for useful suggestions and comments which made for a clearer and stronger manuscript. This work was partially supported by funding from the Center for Exoplanets and Habitable Worlds. The Center for Exoplanets and Habitable Worlds is supported by the Pennsylvania State University, the Eberly College of Science, and the Pennsylvania Space Grant Consortium. This work was supported by NASA Headquarters under the NASA Earth and Space Science Fellowship Program through grants 80NSSC18K1114. We acknowledge support from NSF grants AST-1006676, AST-1126413, AST-1310885, AST-1517592, AST-1310875, AST-1910954, AST-1907622, AST-1909506, ATI 2009889, ATI 2009982, and the NASA Astrobiology Institute (NNA09DA76A) in our pursuit of precision radial velocities in the NIR. We acknowledge support from the Heising-Simons Foundation via grant 2017-0494. Computations for this research were performed on the Pennsylvania State University’s Institute for Computational and Data Sciences’ Roar supercomputer

These results are based on observations obtained with the Habitable-zone Planet Finder Spectrograph on the Hobby-Eberly Telescope. We thank the Resident astronomers and Telescope Operators at the HET for the skillful execution of our observations with HPF, and the HET staff for their dedication to the facility, and enabling these observations. The Hobby-Eberly Telescope is a joint project of the University of Texas at Austin, the Pennsylvania State University, Ludwig-Maximilians-Universität München, and Georg-August Universität Gottingen. The HET is named in honor of its principal benefactors, William P. Hobby and Robert E. Eberly. The HET collaboration acknowledges the support and resources from the Texas Advanced Computing Center. 

This research used the facilities of the Italian Center for Astronomical Archive (IA2) operated by INAF at the Astronomical Observatory of Trieste.

This paper includes data collected by the TESS mission, which are publicly available from the Multimission Archive for Space Telescopes (MAST). Support for MAST for non-HST data is provided by the NASA Office of Space Science via grant NNX09AF08G and by other grants and contracts. This research made use of Lightkurve, a Python package for Kepler and TESS data analysis (Lightkurve Collaboration, 2018).

GS thanks Josh Winn for helpful discussions and conversations.

%

\vspace{5mm}
\facilities{HPF/HET, HARPS-N, TESS.}


\software{
\texttt{astropy} \citep{astropy2013},
\texttt{astroquery} \citep{astroquery},
\texttt{barycorrpy} \citep{kanodia2018}, 
\texttt{batman} \citep{kreidberg2015},
\texttt{dynesty} \citep{speagle2019}, 
\texttt{forecaster} \citep{chen2017}, 
\texttt{GNU Parallel} \citep{Tange2011}, 
\texttt{HxRGproc} \citep{ninan2018},
\texttt{Jupyter} \citep{jupyter2016},
\texttt{juliet} \citep{Espinoza2018},
\texttt{matplotlib} \citep{hunter2007},
\texttt{numpy} \citep{vanderwalt2011},
\texttt{pandas} \citep{pandas2010},
\texttt{radvel} \citep{fulton2018},
\texttt{SERVAL} \citep{zechmeister2018},
\texttt{SOAP 2.0} \citep{dumusque2014},
\texttt{tesscut} \citep{tesscut}.
}



\newpage
\bibliography{ms.bib}

\begin{thebibliography}{}
\expandafter\ifx\csname natexlab\endcsname\relax\def\natexlab#1{#1}\fi
\providecommand{\url}[1]{\href{#1}{#1}}
\providecommand{\dodoi}[1]{doi:~\href{http://doi.org/#1}{\nolinkurl{#1}}}
\providecommand{\doeprint}[1]{\href{http://ascl.net/#1}{\nolinkurl{http://ascl.net/#1}}}
\providecommand{\doarXiv}[1]{\href{https://arxiv.org/abs/#1}{\nolinkurl{https://arxiv.org/abs/#1}}}

\bibitem[{{Anglada-Escud{\'e}} \& {Butler}(2012)}]{anglada2012}
{Anglada-Escud{\'e}}, G., \& {Butler}, R.~P. 2012, \apjs, 200, 15,
  \dodoi{10.1088/0067-0049/200/2/15}

\bibitem[{{Astropy Collaboration} {et~al.}(2013){Astropy Collaboration},
  {Robitaille}, {Tollerud}, {Greenfield}, {Droettboom}, {Bray}, {Aldcroft},
  {Davis}, {Ginsburg}, {Price-Whelan}, {Kerzendorf}, {Conley}, {Crighton},
  {Barbary}, {Muna}, {Ferguson}, {Grollier}, {Parikh}, {Nair}, {Unther},
  {Deil}, {Woillez}, {Conseil}, {Kramer}, {Turner}, {Singer}, {Fox}, {Weaver},
  {Zabalza}, {Edwards}, {Azalee Bostroem}, {Burke}, {Casey}, {Crawford},
  {Dencheva}, {Ely}, {Jenness}, {Labrie}, {Lim}, {Pierfederici}, {Pontzen},
  {Ptak}, {Refsdal}, {Servillat}, \& {Streicher}}]{astropy2013}
{Astropy Collaboration}, {Robitaille}, T.~P., {Tollerud}, E.~J., {et~al.} 2013,
  \aap, 558, A33, \dodoi{10.1051/0004-6361/201322068}

\bibitem[{{Bedell} {et~al.}(2019){Bedell}, {Hogg}, {Foreman-Mackey}, {Montet},
  \& {Luger}}]{bedell2019}
{Bedell}, M., {Hogg}, D.~W., {Foreman-Mackey}, D., {Montet}, B.~T., \& {Luger},
  R. 2019, \aj, 158, 164, \dodoi{10.3847/1538-3881/ab40a7}

\bibitem[{{Boisse} {et~al.}(2011){Boisse}, {Bouchy}, {H{\'e}brard}, {Bonfils},
  {Santos}, \& {Vauclair}}]{boisse2011}
{Boisse}, I., {Bouchy}, F., {H{\'e}brard}, G., {et~al.} 2011, \aap, 528, A4,
  \dodoi{10.1051/0004-6361/201014354}

\bibitem[{{Brasseur} {et~al.}(2019){Brasseur}, {Phillip}, {Fleming},
  {Mullally}, \& {White}}]{tesscut}
{Brasseur}, C.~E., {Phillip}, C., {Fleming}, S.~W., {Mullally}, S.~E., \&
  {White}, R.~L. 2019, {Astrocut: Tools for creating cutouts of TESS images}.
\newblock \doeprint{1905.007}

\bibitem[{{Cauley} {et~al.}(2019){Cauley}, {Shkolnik}, {Llama}, \&
  {Lanza}}]{cauley2019}
{Cauley}, P.~W., {Shkolnik}, E.~L., {Llama}, J., \& {Lanza}, A.~F. 2019, Nature
  Astronomy, 3, 1128, \dodoi{10.1038/s41550-019-0840-x}

\bibitem[{{Chen} \& {Kipping}(2017)}]{chen2017}
{Chen}, J., \& {Kipping}, D. 2017, \apj, 834, 17,
  \dodoi{10.3847/1538-4357/834/1/17}

\bibitem[{{Cosentino} {et~al.}(2012){Cosentino}, {Lovis}, {Pepe}, {Collier
  Cameron}, {Latham}, {Molinari}, {Udry}, {Bezawada}, {Black}, {Born},
  {Buchschacher}, {Charbonneau}, {Figueira}, {Fleury}, {Galli}, {Gallie},
  {Gao}, {Ghedina}, {Gonzalez}, {Gonzalez}, {Guerra}, {Henry}, {Horne},
  {Hughes}, {Kelly}, {Lodi}, {Lunney}, {Maire}, {Mayor}, {Micela}, {Ordway},
  {Peacock}, {Phillips}, {Piotto}, {Pollacco}, {Queloz}, {Rice}, {Riverol},
  {Riverol}, {San Juan}, {Sasselov}, {Segransan}, {Sozzetti}, {Sosnowska},
  {Stobie}, {Szentgyorgyi}, {Vick}, \& {Weber}}]{cosentino2012}
{Cosentino}, R., {Lovis}, C., {Pepe}, F., {et~al.} 2012, in Society of
  Photo-Optical Instrumentation Engineers (SPIE) Conference Series, Vol. 8446,
  \procspie, 84461V, \dodoi{10.1117/12.925738}

\bibitem[{{Dawson} \& {Fabrycky}(2010)}]{dawson2010}
{Dawson}, R.~I., \& {Fabrycky}, D.~C. 2010, \apj, 722, 937,
  \dodoi{10.1088/0004-637X/722/1/937}

\bibitem[{{Dumusque} {et~al.}(2014){Dumusque}, {Boisse}, \&
  {Santos}}]{dumusque2014}
{Dumusque}, X., {Boisse}, I., \& {Santos}, N.~C. 2014, \apj, 796, 132,
  \dodoi{10.1088/0004-637X/796/2/132}

\bibitem[{{Engle} \& {Guinan}(2011)}]{engle2011}
{Engle}, S.~G., \& {Guinan}, E.~F. 2011, in Astronomical Society of the Pacific
  Conference Series, Vol. 451, 9th Pacific Rim Conference on Stellar
  Astrophysics, ed. S.~{Qain}, K.~{Leung}, L.~{Zhu}, \& S.~{Kwok}, 285.
\newblock \doarXiv{1111.2872}

\bibitem[{{Espinoza} {et~al.}(2018){Espinoza}, {Kossakowski}, \&
  {Brahm}}]{Espinoza2018}
{Espinoza}, N., {Kossakowski}, D., \& {Brahm}, R. 2018, arXiv e-prints,
  arXiv:1812.08549.
\newblock \doarXiv{1812.08549}

\bibitem[{{Feng} {et~al.}(2017){Feng}, {Tuomi}, \& {Jones}}]{feng2017}
{Feng}, F., {Tuomi}, M., \& {Jones}, H.~R.~A. 2017, \mnras, 470, 4794,
  \dodoi{10.1093/mnras/stx1126}

\bibitem[{{Fischer} \& {Saur}(2019)}]{fischer2019}
{Fischer}, C., \& {Saur}, J. 2019, \apj, 872, 113,
  \dodoi{10.3847/1538-4357/aafaf2}

\bibitem[{{Ford} \& {Gregory}(2007)}]{ford2007}
{Ford}, E.~B., \& {Gregory}, P.~C. 2007, in Astronomical Society of the Pacific
  Conference Series, Vol. 371, Statistical Challenges in Modern Astronomy IV,
  ed. G.~J. {Babu} \& E.~D. {Feigelson}, 189.
\newblock \doarXiv{astro-ph/0608328}

\bibitem[{{Foster} {et~al.}(2020){Foster}, {Poppenhaeger},
  {Alvarado-G{\'o}mez}, \& {Schmitt}}]{foster2020}
{Foster}, G., {Poppenhaeger}, K., {Alvarado-G{\'o}mez}, J.~D., \& {Schmitt},
  J.~H.~M.~M. 2020, \mnras, 497, 1015, \dodoi{10.1093/mnras/staa1982}

\bibitem[{{Fulton} {et~al.}(2018){Fulton}, {Petigura}, {Blunt}, \&
  {Sinukoff}}]{fulton2018}
{Fulton}, B.~J., {Petigura}, E.~A., {Blunt}, S., \& {Sinukoff}, E. 2018, \pasp,
  130, 044504, \dodoi{10.1088/1538-3873/aaaaa8}

\bibitem[{{Gillon} {et~al.}(2017){Gillon}, {Triaud}, {Demory}, {Jehin}, {Agol},
  {Deck}, {Lederer}, {de Wit}, {Burdanov}, {Ingalls}, {Bolmont}, {Leconte},
  {Raymond}, {Selsis}, {Turbet}, {Barkaoui}, {Burgasser}, {Burleigh}, {Carey},
  {Chaushev}, {Copperwheat}, {Delrez}, {Fernand es}, {Holdsworth}, {Kotze},
  {Van Grootel}, {Almleaky}, {Benkhaldoun}, {Magain}, \& {Queloz}}]{gillon2017}
{Gillon}, M., {Triaud}, A. H.~M.~J., {Demory}, B.-O., {et~al.} 2017, \nat, 542,
  456, \dodoi{10.1038/nature21360}

\bibitem[{Ginsburg {et~al.}(2018)Ginsburg, Sipocz, Parikh, Woillez, Groener,
  Liedtke, Robitaille, Deil, jcsegovia, Norman, Svoboda, Brasseur, Tollerud,
  Persson, adamginsburg, Séguin-Charbonneau, Armstrong, de~Val-Borro, Morris,
  Mirocha, Yadav, Seifert, Droettboom, Moolekamp, james allen, Bostroem,
  Egeland, Singer, Rol, \& Grollier}]{astroquery}
Ginsburg, A., Sipocz, B., Parikh, M., {et~al.} 2018, astropy/astroquery: v0.3.7
  release, \dodoi{10.5281/zenodo.1160627}

\bibitem[{{Gomes da Silva} {et~al.}(2011){Gomes da Silva}, {Santos}, {Bonfils},
  {Delfosse}, {Forveille}, \& {Udry}}]{gds2011}
{Gomes da Silva}, J., {Santos}, N.~C., {Bonfils}, X., {et~al.} 2011, \aap, 534,
  A30, \dodoi{10.1051/0004-6361/201116971}

\bibitem[{{Hallinan} {et~al.}(2015){Hallinan}, {Littlefair}, {Cotter},
  {Bourke}, {Harding}, {Pineda}, {Butler}, {Golden}, {Basri}, {Doyle}, {Kao},
  {Berdyugina}, {Kuznetsov}, {Rupen}, \& {Antonova}}]{hallinan2015}
{Hallinan}, G., {Littlefair}, S.~P., {Cotter}, G., {et~al.} 2015, \nat, 523,
  568, \dodoi{10.1038/nature14619}

\bibitem[{{Hardegree-Ullman} {et~al.}(2019){Hardegree-Ullman}, {Cushing},
  {Muirhead}, \& {Christiansen}}]{hardegree2019}
{Hardegree-Ullman}, K.~K., {Cushing}, M.~C., {Muirhead}, P.~S., \&
  {Christiansen}, J.~L. 2019, \aj, 158, 75, \dodoi{10.3847/1538-3881/ab21d2}

\bibitem[{{Huang} {et~al.}(2018){Huang}, {Burt}, {Vanderburg}, {G{\"u}nther},
  {Shporer}, {Dittmann}, {Winn}, {Wittenmyer}, {Sha}, {Kane}, {Ricker}, {Vand
  erspek}, {Latham}, {Seager}, {Jenkins}, {Caldwell}, {Collins}, {Guerrero},
  {Smith}, {Quinn}, {Udry}, {Pepe}, {Bouchy}, {S{\'e}gransan}, {Lovis},
  {Ehrenreich}, {Marmier}, {Mayor}, {Wohler}, {Haworth}, {Morgan}, {Fausnaugh},
  {Ciardi}, {Christiansen}, {Charbonneau}, {Dragomir}, {Deming}, {Glidden},
  {Levine}, {McCullough}, {Yu}, {Narita}, {Nguyen}, {Morton}, {Pepper},
  {P{\'a}l}, {Rodriguez}, {Stassun}, {Torres}, {Sozzetti}, {Doty},
  {Christensen-Dalsgaard}, {Laughlin}, {Clampin}, {Bean}, {Buchhave}, {Bakos},
  {Sato}, {Ida}, {Kaltenegger}, {Palle}, {Sasselov}, {Butler}, {Lissauer},
  {Ge}, \& {Rinehart}}]{huang2018}
{Huang}, C.~X., {Burt}, J., {Vanderburg}, A., {et~al.} 2018, \apjl, 868, L39,
  \dodoi{10.3847/2041-8213/aaef91}

\bibitem[{{Hunter}(2007)}]{hunter2007}
{Hunter}, J.~D. 2007, Computing in Science and Engineering, 9, 90,
  \dodoi{10.1109/MCSE.2007.55}

\bibitem[{Jenkins {et~al.}(2016)Jenkins, Twicken, McCauliff, Campbell,
  Sanderfer, Lung, Mansouri-Samani, Girouard, Tenenbaum, Klaus, Smith,
  Caldwell, Chacon, Henze, Heiges, Latham, Morgan, Swade, Rinehart, \&
  Vanderspek}]{jenkins2016}
Jenkins, J.~M., Twicken, J.~D., McCauliff, S., {et~al.} 2016, in Software and
  Cyberinfrastructure for Astronomy IV, ed. G.~Chiozzi \& J.~C. Guzman, Vol.
  9913, International Society for Optics and Photonics (SPIE), 1232 -- 1251,
  \dodoi{10.1117/12.2233418}

\bibitem[{{Kanodia} \& {Wright}(2018)}]{kanodia2018}
{Kanodia}, S., \& {Wright}, J. 2018, Research Notes of the American
  Astronomical Society, 2, 4, \dodoi{10.3847/2515-5172/aaa4b7}

\bibitem[{{Kaplan} {et~al.}(2019){Kaplan}, {Bender}, {Terrien}, {Ninan}, {Roy},
  \& {Mahadevan}}]{kaplan2019}
{Kaplan}, K.~F., {Bender}, C.~F., {Terrien}, R.~C., {et~al.} 2019, in
  Astronomical Society of the Pacific Conference Series, Vol. 523, Astronomical
  Data Analysis Software and Systems XXVII, ed. P.~J. {Teuben}, M.~W. {Pound},
  B.~A. {Thomas}, \& E.~M. {Warner}, 567

\bibitem[{Kass \& Raftery(1995)}]{kass1995}
Kass, R.~E., \& Raftery, A.~E. 1995, Journal of the American Statistical
  Association, 90, 773, \dodoi{10.1080/01621459.1995.10476572}

\bibitem[{Kluyver {et~al.}(2016)Kluyver, Ragan-Kelley, P{\'e}rez, Granger,
  Bussonnier, Frederic, Kelley, Hamrick, Grout, Corlay, Ivanov, Avila, Abdalla,
  Willing, \& development team}]{jupyter2016}
Kluyver, T., Ragan-Kelley, B., P{\'e}rez, F., {et~al.} 2016, in Positioning and
  Power in Academic Publishing: Players, Agents and Agendas, ed. F.~Loizides \&
  B.~Scmidt (IOS Press), 87--90.
\newblock \url{https://eprints.soton.ac.uk/403913/}

\bibitem[{{Kov{\'a}cs} {et~al.}(2002){Kov{\'a}cs}, {Zucker}, \&
  {Mazeh}}]{kovacs2002}
{Kov{\'a}cs}, G., {Zucker}, S., \& {Mazeh}, T. 2002, \aap, 391, 369,
  \dodoi{10.1051/0004-6361:20020802}

\bibitem[{{Kreidberg}(2015)}]{kreidberg2015}
{Kreidberg}, L. 2015, \pasp, 127, 1161, \dodoi{10.1086/683602}

\bibitem[{{K{\"u}rster} {et~al.}(2003){K{\"u}rster}, {Endl}, {Rouesnel}, {Els},
  {Kaufer}, {Brillant}, {Hatzes}, {Saar}, \& {Cochran}}]{kurster2003}
{K{\"u}rster}, M., {Endl}, M., {Rouesnel}, F., {et~al.} 2003, \aap, 403, 1077,
  \dodoi{10.1051/0004-6361:20030396}

\bibitem[{{Lanza}(2009)}]{lanza2009}
{Lanza}, A.~F. 2009, \aap, 505, 339, \dodoi{10.1051/0004-6361/200912367}

\bibitem[{{Lightkurve Collaboration} {et~al.}(2018){Lightkurve Collaboration},
  {Cardoso}, {Hedges}, {Gully-Santiago}, {Saunders}, {Cody}, {Barclay}, {Hall},
  {Sagear}, {Turtelboom}, {Zhang}, {Tzanidakis}, {Mighell}, {Coughlin}, {Bell},
  {Berta-Thompson}, {Williams}, {Dotson}, \& {Barentsen}}]{lightkurve}
{Lightkurve Collaboration}, {Cardoso}, J.~V.~d.~M., {Hedges}, C., {et~al.}
  2018, {Lightkurve: Kepler and TESS time series analysis in Python},
  Astrophysics Source Code Library.
\newblock \doeprint{1812.013}

\bibitem[{{Littlefair} {et~al.}(2008){Littlefair}, {Dhillon}, {Marsh},
  {Shahbaz}, {Mart{\'\i}n}, \& {Copperwheat}}]{littlefair2008}
{Littlefair}, S.~P., {Dhillon}, V.~S., {Marsh}, T.~R., {et~al.} 2008, \mnras,
  391, L88, \dodoi{10.1111/j.1745-3933.2008.00562.x}

\bibitem[{{Mahadevan} {et~al.}(2012){Mahadevan}, {Ramsey}, {Bender}, {Terrien},
  {Wright}, {Halverson}, {Hearty}, {Nelson}, {Burton}, {Redman}, {Osterman},
  {Diddams}, {Kasting}, {Endl}, \& {Deshpande}}]{mahadevan2012}
{Mahadevan}, S., {Ramsey}, L., {Bender}, C., {et~al.} 2012, in \procspie, Vol.
  8446, Ground-based and Airborne Instrumentation for Astronomy IV, 84461S,
  \dodoi{10.1117/12.926102}

\bibitem[{{Mahadevan} {et~al.}(2014){Mahadevan}, {Ramsey}, {Terrien},
  {Halverson}, {Roy}, {Hearty}, {Levi}, {Stefansson}, {Robertson}, {Bender},
  {Schwab}, \& {Nelson}}]{mahadevan2014}
{Mahadevan}, S., {Ramsey}, L.~W., {Terrien}, R., {et~al.} 2014, in Society of
  Photo-Optical Instrumentation Engineers (SPIE) Conference Series, Vol. 9147,
  Ground-based and Airborne Instrumentation for Astronomy V, 91471G,
  \dodoi{10.1117/12.2056417}

\bibitem[{{Marchwinski} {et~al.}(2015){Marchwinski}, {Mahadevan}, {Robertson},
  {Ramsey}, \& {Harder}}]{marchwinski2015}
{Marchwinski}, R.~C., {Mahadevan}, S., {Robertson}, P., {Ramsey}, L., \&
  {Harder}, J. 2015, \apj, 798, 63, \dodoi{10.1088/0004-637X/798/1/63}

\bibitem[{McKinney(2010)}]{pandas2010}
McKinney, W. 2010, in Proceedings of the 9th Python in Science Conference, ed.
  S.~van~der Walt \& J.~Millman, 51 -- 56

\bibitem[{Metcalf {et~al.}(2019)Metcalf, Anderson, Bender, Blakeslee, Brand,
  Carlson, Cochran, Diddams, Endl, Fredrick, Halverson, Hickstein, Hearty,
  Jennings, Kanodia, Kaplan, Levi, Lubar, Mahadevan, Monson, Ninan, Nitroy,
  Osterman, Papp, Quinlan, Ramsey, Robertson, Roy, Schwab, Sigurdsson,
  Srinivasan, Stefansson, Sterner, Terrien, Wolszczan, Wright, \&
  Ycas}]{metcalf2019}
Metcalf, A.~J., Anderson, T., Bender, C.~F., {et~al.} 2019, Optica, 6, 233,
  \dodoi{10.1364/OPTICA.6.000233}

\bibitem[{{Mortier} {et~al.}(2015){Mortier}, {Faria}, {Correia}, {Santerne}, \&
  {Santos}}]{mortier2015}
{Mortier}, A., {Faria}, J.~P., {Correia}, C.~M., {Santerne}, A., \& {Santos},
  N.~C. 2015, \aap, 573, A101, \dodoi{10.1051/0004-6361/201424908}

\bibitem[{{Neubauer}(1980)}]{neubauer1980}
{Neubauer}, F.~M. 1980, \jgr, 85, 1171, \dodoi{10.1029/JA085iA03p01171}

\bibitem[{{Newton} {et~al.}(2016){Newton}, {Irwin}, {Charbonneau},
  {Berta-Thompson}, {Dittmann}, \& {West}}]{newton2016}
{Newton}, E.~R., {Irwin}, J., {Charbonneau}, D., {et~al.} 2016, \apj, 821, 93,
  \dodoi{10.3847/0004-637X/821/2/93}

\bibitem[{{Ninan} {et~al.}(2018){Ninan}, {Bender}, {Mahadevan}, {Ford},
  {Monson}, {Kaplan}, {Terrien}, {Roy}, {Robertson}, {Kanodia}, \&
  {Stefansson}}]{ninan2018}
{Ninan}, J.~P., {Bender}, C.~F., {Mahadevan}, S., {et~al.} 2018, in Society of
  Photo-Optical Instrumentation Engineers (SPIE) Conference Series, Vol. 10709,
  High Energy, Optical, and Infrared Detectors for Astronomy VIII, 107092U,
  \dodoi{10.1117/12.2312787}

\bibitem[{{Parker}(1958)}]{parker1958}
{Parker}, E.~N. 1958, \apj, 128, 664, \dodoi{10.1086/146579}

\bibitem[{{P{\'e}rez-Torres} {et~al.}(2021){P{\'e}rez-Torres}, {G{\'o}mez},
  {Ortiz}, {Leto}, {Anglada}, {G{\'o}mez}, {Rodr{\'\i}guez}, {Trigilio},
  {Amado}, {Alberdi}, {Anglada-Escud{\'e}}, {Osorio}, {Umana}, {Berdi{\~n}as},
  {L{\'o}pez-Gonz{\'a}lez}, {Morales}, {Rodr{\'\i}guez-L{\'o}pez}, \&
  {Chibueze}}]{perez2021}
{P{\'e}rez-Torres}, M., {G{\'o}mez}, J.~F., {Ortiz}, J.~L., {et~al.} 2021,
  \aap, 645, A77, \dodoi{10.1051/0004-6361/202039052}

\bibitem[{{Pineda} \& {Hallinan}(2018)}]{pineda2018}
{Pineda}, J.~S., \& {Hallinan}, G. 2018, \apj, 866, 155,
  \dodoi{10.3847/1538-4357/aae078}

\bibitem[{{Pope} {et~al.}(2020){Pope}, {Bedell}, {Callingham}, {Vedantham},
  {Snellen}, {Price-Whelan}, \& {Shimwell}}]{pope2020}
{Pope}, B. J.~S., {Bedell}, M., {Callingham}, J.~R., {et~al.} 2020, \apjl, 890,
  L19, \dodoi{10.3847/2041-8213/ab5b99}

\bibitem[{{Rasio} \& {Ford}(1996)}]{rasio1996}
{Rasio}, F.~A., \& {Ford}, E.~B. 1996, Science, 274, 954,
  \dodoi{10.1126/science.274.5289.954}

\bibitem[{{Reiners} {et~al.}(2010){Reiners}, {Bean}, {Huber}, {Dreizler},
  {Seifahrt}, \& {Czesla}}]{reiners2010}
{Reiners}, A., {Bean}, J.~L., {Huber}, K.~F., {et~al.} 2010, \apj, 710, 432,
  \dodoi{10.1088/0004-637X/710/1/432}

\bibitem[{{Reiners} {et~al.}(2018){Reiners}, {Zechmeister}, \&
  {Caballero}}]{reiners2018}
{Reiners}, A., {Zechmeister}, M., \& {Caballero}, J.~A. 2018, \aap, 612, A49,
  \dodoi{10.1051/0004-6361/201732054}

\bibitem[{{Ricker} {et~al.}(2015){Ricker}, {Winn}, {Vanderspek}, {Latham},
  {Bakos}, {Bean}, {Berta-Thompson}, {Brown}, {Buchhave}, {Butler}, {Butler},
  {Chaplin}, {Charbonneau}, {Christensen-Dalsgaard}, {Clampin}, {Deming},
  {Doty}, {De Lee}, {Dressing}, {Dunham}, {Endl}, {Fressin}, {Ge}, {Henning},
  {Holman}, {Howard}, {Ida}, {Jenkins}, {Jernigan}, {Johnson}, {Kaltenegger},
  {Kawai}, {Kjeldsen}, {Laughlin}, {Levine}, {Lin}, {Lissauer}, {MacQueen},
  {Marcy}, {McCullough}, {Morton}, {Narita}, {Paegert}, {Palle}, {Pepe},
  {Pepper}, {Quirrenbach}, {Rinehart}, {Sasselov}, {Sato}, {Seager},
  {Sozzetti}, {Stassun}, {Sullivan}, {Szentgyorgyi}, {Torres}, {Udry}, \&
  {Villasenor}}]{ricker2015}
{Ricker}, G.~R., {Winn}, J.~N., {Vanderspek}, R., {et~al.} 2015, Journal of
  Astronomical Telescopes, Instruments, and Systems, 1, 014003,
  \dodoi{10.1117/1.JATIS.1.1.014003}

\bibitem[{{Robertson}(2018)}]{robertson2018}
{Robertson}, P. 2018, \apjl, 864, L28, \dodoi{10.3847/2041-8213/aadc0b}

\bibitem[{{Robertson} {et~al.}(2013){Robertson}, {Endl}, {Cochran}, \&
  {Dodson-Robinson}}]{robertson2013}
{Robertson}, P., {Endl}, M., {Cochran}, W.~D., \& {Dodson-Robinson}, S.~E.
  2013, \apj, 764, 3, \dodoi{10.1088/0004-637X/764/1/3}

\bibitem[{{Robertson} {et~al.}(2014){Robertson}, {Mahadevan}, {Endl}, \&
  {Roy}}]{robertson2014}
{Robertson}, P., {Mahadevan}, S., {Endl}, M., \& {Roy}, A. 2014, Science, 345,
  440, \dodoi{10.1126/science.1253253}

\bibitem[{{Saur} {et~al.}(2013){Saur}, {Grambusch}, {Duling}, {Neubauer}, \&
  {Simon}}]{saur2013}
{Saur}, J., {Grambusch}, T., {Duling}, S., {Neubauer}, F.~M., \& {Simon}, S.
  2013, \aap, 552, A119, \dodoi{10.1051/0004-6361/201118179}

\bibitem[{{Shetrone} {et~al.}(2007){Shetrone}, {Cornell}, {Fowler}, {Gaffney},
  {Laws}, {Mader}, {Mason}, {Odewahn}, {Roman}, {Rostopchin}, {Schneider},
  {Umbarger}, \& {Westfall}}]{shetrone2007}
{Shetrone}, M., {Cornell}, M.~E., {Fowler}, J.~R., {et~al.} 2007, \pasp, 119,
  556, \dodoi{10.1086/519291}

\bibitem[{{Shkolnik} {et~al.}(2008){Shkolnik}, {Bohlender}, {Walker}, \&
  {Collier Cameron}}]{shkolnik2008}
{Shkolnik}, E., {Bohlender}, D.~A., {Walker}, G. A.~H., \& {Collier Cameron},
  A. 2008, \apj, 676, 628, \dodoi{10.1086/527351}

\bibitem[{{Shkolnik} {et~al.}(2005){Shkolnik}, {Walker}, {Bohlender}, {Gu}, \&
  {K{\"u}rster}}]{shkolnik2005}
{Shkolnik}, E., {Walker}, G.~A.~H., {Bohlender}, D.~A., {Gu}, P.~G., \&
  {K{\"u}rster}, M. 2005, \apj, 622, 1075, \dodoi{10.1086/428037}

\bibitem[{{Shporer}(2017)}]{shporer2017}
{Shporer}, A. 2017, \pasp, 129, 072001, \dodoi{10.1088/1538-3873/aa7112}

\bibitem[{Smith {et~al.}(2012)Smith, Stumpe, Cleve, Jenkins, Barclay, Fanelli,
  Girouard, Kolodziejczak, McCauliff, Morris, \& Twicken}]{smith2012}
Smith, J.~C., Stumpe, M.~C., Cleve, J. E.~V., {et~al.} 2012, Publications of
  the Astronomical Society of the Pacific, 124, 1000, \dodoi{10.1086/667697}

\bibitem[{{Speagle}(2019)}]{speagle2019}
{Speagle}, J.~S. 2019, arXiv e-prints, arXiv:1904.02180.
\newblock \doarXiv{1904.02180}

\bibitem[{{Stassun} {et~al.}(2018){Stassun}, {Oelkers}, {Pepper}, {Paegert},
  {De Lee}, {Torres}, {Latham}, {Charpinet}, {Dressing}, {Huber}, {Kane},
  {L{\'e}pine}, {Mann}, {Muirhead}, {Rojas-Ayala}, {Silvotti}, {Fleming},
  {Levine}, \& {Plavchan}}]{stassun2018}
{Stassun}, K.~G., {Oelkers}, R.~J., {Pepper}, J., {et~al.} 2018, \aj, 156, 102,
  \dodoi{10.3847/1538-3881/aad050}

\bibitem[{{Stassun} {et~al.}(2019){Stassun}, {Oelkers}, {Paegert}, {Torres},
  {Pepper}, {De Lee}, {Collins}, {Latham}, {Muirhead}, {Chittidi},
  {Rojas-Ayala}, {Fleming}, {Rose}, {Tenenbaum}, {Ting}, {Kane}, {Barclay},
  {Bean}, {Brassuer}, {Charbonneau}, {Ge}, {Lissauer}, {Mann}, {McLean},
  {Mullally}, {Narita}, {Plavchan}, {Ricker}, {Sasselov}, {Seager}, {Sharma},
  {Shiao}, {Sozzetti}, {Stello}, {Vanderspek}, {Wallace}, \&
  {Winn}}]{stassun2019}
{Stassun}, K.~G., {Oelkers}, R.~J., {Paegert}, M., {et~al.} 2019, \aj, 158,
  138, \dodoi{10.3847/1538-3881/ab3467}

\bibitem[{{Stefansson} {et~al.}(2016){Stefansson}, {Hearty}, {Robertson},
  {Mahadevan}, {Anderson}, {Levi}, {Bender}, {Nelson}, {Monson}, {Blank},
  {Halverson}, {Henderson}, {Ramsey}, {Roy}, {Schwab}, \&
  {Terrien}}]{stefansson2016}
{Stefansson}, G., {Hearty}, F., {Robertson}, P., {et~al.} 2016, \apj, 833, 175,
  \dodoi{10.3847/1538-4357/833/2/175}

\bibitem[{{Stefansson} {et~al.}(2020){Stefansson}, {Ca{\~n}as}, {Wisniewski},
  {Robertson}, {Mahadevan}, {Maney}, {Kanodia}, {Beard}, {Bender}, {Brunt},
  {Clemens}, {Cochran}, {Diddams}, {Endl}, {Ford}, {Fredrick}, {Halverson},
  {Hearty}, {Hebb}, {Huehnerhoff}, {Jennings}, {Kaplan}, {Levi}, {Lubar},
  {Metcalf}, {Monson}, {Morris}, {Ninan}, {Nitroy}, {Ramsey}, {Roy}, {Schwab},
  {Sigurdsson}, {Terrien}, \& {Wright}}]{stefansson2020}
{Stefansson}, G., {Ca{\~n}as}, C., {Wisniewski}, J., {et~al.} 2020, \aj, 159,
  100, \dodoi{10.3847/1538-3881/ab5f15}

\bibitem[{Stumpe {et~al.}(2014)Stumpe, Smith, Catanzarite, Cleve, Jenkins,
  Twicken, \& Girouard}]{stumpe2014}
Stumpe, M.~C., Smith, J.~C., Catanzarite, J.~H., {et~al.} 2014, Publications of
  the Astronomical Society of the Pacific, 126, 100, \dodoi{10.1086/674989}

\bibitem[{Tange(2011)}]{Tange2011}
Tange, O. 2011, ;login: The USENIX Magazine, 36, 42,
  \dodoi{10.5281/zenodo.16303}

\bibitem[{{Tenenbaum} \& {Jenkins}(2018)}]{tenenbaum2018}
{Tenenbaum}, P., \& {Jenkins}, J. 2018, TESS Science Data Products Description
  Document, EXP-TESS-ARC-ICD-0014 Rev D.
\newblock
  \url{https://archive.stsci.edu/missions/tess/doc/EXP-TESS-ARC-ICD-TM-0014.pdf}

\bibitem[{{Turner} {et~al.}(2020){Turner}, {Zarka}, {Grie{\ss}meier}, {Lazio},
  {Cecconi}, {Enriquez}, {Girard}, {Jayawardhana}, {Lamy}, {Nichols}, \& {de
  Pater}}]{turner2020}
{Turner}, J.~D., {Zarka}, P., {Grie{\ss}meier}, J.-M., {et~al.} 2020, arXiv
  e-prints, arXiv:2012.07926.
\newblock \doarXiv{2012.07926}

\bibitem[{{Turnpenney} {et~al.}(2018){Turnpenney}, {Nichols}, {Wynn}, \&
  {Burleigh}}]{turnpenney2018}
{Turnpenney}, S., {Nichols}, J.~D., {Wynn}, G.~A., \& {Burleigh}, M.~R. 2018,
  \apj, 854, 72, \dodoi{10.3847/1538-4357/aaa59c}

\bibitem[{{Van Der Walt} {et~al.}(2011){Van Der Walt}, {Colbert}, \&
  {Varoquaux}}]{vanderwalt2011}
{Van Der Walt}, S., {Colbert}, S.~C., \& {Varoquaux}, G. 2011, ArXiv e-prints.
\newblock \doarXiv{1102.1523}

\bibitem[{{van Haarlem} {et~al.}(2013){van Haarlem}, {Wise, M. W.}, {Gunst, A.
  W.}, {Heald, G.}, {McKean, J. P.}, {Hessels, J. W. T.}, {de Bruyn, A. G.},
  {Nijboer, R.}, {Swinbank, J.}, {Fallows, R.}, {Brentjens, M.}, {Nelles, A.},
  {Beck, R.}, {Falcke, H.}, {Fender, R.}, {H\"orandel, J.}, {Koopmans, L. V.
  E.}, {Mann, G.}, {Miley, G.}, {R\"ottgering, H.}, {Stappers, B. W.}, {Wijers,
  R. A. M. J.}, {Zaroubi, S.}, {van den Akker, M.}, {Alexov, A.}, {Anderson,
  J.}, {Anderson, K.}, {van Ardenne, A.}, {Arts, M.}, {Asgekar, A.}, {Avruch,
  I. M.}, {Batejat, F.}, {B\"ahren, L.}, {Bell, M. E.}, {Bell, M. R.}, {van
  Bemmel, I.}, {Bennema, P.}, {Bentum, M. J.}, {Bernardi, G.}, {Best, P.},
  {B\^{\i}rzan, L.}, {Bonafede, A.}, {Boonstra, A.-J.}, {Braun, R.}, {Bregman,
  J.}, {Breitling, F.}, {van de Brink, R. H.}, {Broderick, J.}, {Broekema, P.
  C.}, {Brouw, W. N.}, {Br\"uggen, M.}, {Butcher, H. R.}, {van Cappellen, W.},
  {Ciardi, B.}, {Coenen, T.}, {Conway, J.}, {Coolen, A.}, {Corstanje, A.},
  {Damstra, S.}, {Davies, O.}, {Deller, A. T.}, {Dettmar, R.-J.}, {van Diepen,
  G.}, {Dijkstra, K.}, {Donker, P.}, {Doorduin, A.}, {Dromer, J.}, {Drost, M.},
  {van Duin, A.}, {Eisl\"offel, J.}, {van Enst, J.}, {Ferrari, C.}, {Frieswijk,
  W.}, {Gankema, H.}, {Garrett, M. A.}, {de Gasperin, F.}, {Gerbers, M.}, {de
  Geus, E.}, {Grie\ss{}meier, J.-M.}, {Grit, T.}, {Gruppen, P.}, {Hamaker, J.
  P.}, {Hassall, T.}, {Hoeft, M.}, {Holties, H. A.}, {Horneffer, A.}, {van der
  Horst, A.}, {van Houwelingen, A.}, {Huijgen, A.}, {Iacobelli, M.}, {Intema,
  H.}, {Jackson, N.}, {Jelic, V.}, {de Jong, A.}, {Juette, E.}, {Kant, D.},
  {Karastergiou, A.}, {Koers, A.}, {Kollen, H.}, {Kondratiev, V. I.},
  {Kooistra, E.}, {Koopman, Y.}, {Koster, A.}, {Kuniyoshi, M.}, {Kramer, M.},
  {Kuper, G.}, {Lambropoulos, P.}, {Law, C.}, {van Leeuwen, J.}, {Lemaitre,
  J.}, {Loose, M.}, {Maat, P.}, {Macario, G.}, {Markoff, S.}, {Masters, J.},
  {McFadden, R. A.}, {McKay-Bukowski, D.}, {Meijering, H.}, {Meulman, H.},
  {Mevius, M.}, {Middelberg, E.}, {Millenaar, R.}, {Miller-Jones, J. C. A.},
  {Mohan, R. N.}, {Mol, J. D.}, {Morawietz, J.}, {Morganti, R.}, {Mulcahy, D.
  D.}, {Mulder, E.}, {Munk, H.}, {Nieuwenhuis, L.}, {van Nieuwpoort, R.},
  {Noordam, J. E.}, {Norden, M.}, {Noutsos, A.}, {Offringa, A. R.}, {Olofsson,
  H.}, {Omar, A.}, {Orr\'u, E.}, {Overeem, R.}, {Paas, H.}, {Pandey-Pommier,
  M.}, {Pandey, V. N.}, {Pizzo, R.}, {Polatidis, A.}, {Rafferty, D.},
  {Rawlings, S.}, {Reich, W.}, {de Reijer, J.-P.}, {Reitsma, J.}, {Renting, G.
  A.}, {Riemers, P.}, {Rol, E.}, {Romein, J. W.}, {Roosjen, J.}, {Ruiter, M.},
  {Scaife, A.}, {van der Schaaf, K.}, {Scheers, B.}, {Schellart, P.},
  {Schoenmakers, A.}, {Schoonderbeek, G.}, {Serylak, M.}, {Shulevski, A.},
  {Sluman, J.}, {Smirnov, O.}, {Sobey, C.}, {Spreeuw, H.}, {Steinmetz, M.},
  {Sterks, C. G. M.}, {Stiepel, H.-J.}, {Stuurwold, K.}, {Tagger, M.}, {Tang,
  Y.}, {Tasse, C.}, {Thomas, I.}, {Thoudam, S.}, {Toribio, M. C.}, {van der
  Tol, B.}, {Usov, O.}, {van Veelen, M.}, {van der Veen, A.-J.}, {ter Veen,
  S.}, {Verbiest, J. P. W.}, {Vermeulen, R.}, {Vermaas, N.}, {Vocks, C.},
  {Vogt, C.}, {de Vos, M.}, {van der Wal, E.}, {van Weeren, R.}, {Weggemans,
  H.}, {Weltevrede, P.}, {White, S.}, {Wijnholds, S. J.}, {Wilhelmsson, T.},
  {Wucknitz, O.}, {Yatawatta, S.}, {Zarka, P.}, {Zensus, A.}, \& {van Zwieten,
  J.}}]{haarlem2013}
{van Haarlem}, M.~P., {Wise, M. W.}, {Gunst, A. W.}, {et~al.} 2013, A\&A, 556,
  A2, \dodoi{10.1051/0004-6361/201220873}

\bibitem[{{Vaughan} {et~al.}(1978){Vaughan}, {Preston}, \&
  {Wilson}}]{vaughan1978}
{Vaughan}, A.~H., {Preston}, G.~W., \& {Wilson}, O.~C. 1978, \pasp, 90, 267,
  \dodoi{10.1086/130324}

\bibitem[{{Vedantham} {et~al.}(2020){Vedantham}, {Callingham}, {Shimwell},
  {Tasse}, {Pope}, {Bedell}, {Snellen}, {Best}, {Hardcastle}, {Haverkorn},
  {Mechev}, {O'Sullivan}, {R{\"o}ttgering}, \& {White}}]{vedantham2020}
{Vedantham}, H.~K., {Callingham}, J.~R., {Shimwell}, T.~W., {et~al.} 2020,
  Nature Astronomy, \dodoi{10.1038/s41550-020-1011-9}

\bibitem[{{Zarka} {et~al.}(2015){Zarka}, {Lazio}, \& {Hallinan}}]{zarka2015}
{Zarka}, P., {Lazio}, J., \& {Hallinan}, G. 2015, in Advancing Astrophysics
  with the Square Kilometre Array (AASKA14), 120

\bibitem[{{Zechmeister} \& {K{\"u}rster}(2009)}]{zk09}
{Zechmeister}, M., \& {K{\"u}rster}, M. 2009, \aap, 496, 577,
  \dodoi{10.1051/0004-6361:200811296}

\bibitem[{{Zechmeister} {et~al.}(2018){Zechmeister}, {Reiners}, {Amado},
  {Azzaro}, {Bauer}, {B{\'e}jar}, {Caballero}, {Guenther}, {Hagen}, {Jeffers},
  {Kaminski}, {K{\"u}rster}, {Launhardt}, {Montes}, {Morales}, {Quirrenbach},
  {Reffert}, {Ribas}, {Seifert}, {Tal-Or}, \& {Wolthoff}}]{zechmeister2018}
{Zechmeister}, M., {Reiners}, A., {Amado}, P.~J., {et~al.} 2018, \aap, 609,
  A12, \dodoi{10.1051/0004-6361/201731483}

\bibitem[{{Zic} {et~al.}(2020){Zic}, {Murphy}, {Lynch}, {Heald}, {Lenc},
  {Kaplan}, {Cairns}, {Coward}, {Gendre}, {Johnston}, {MacGregor}, {Price}, \&
  {Wheatland}}]{zic2020}
{Zic}, A., {Murphy}, T., {Lynch}, C., {et~al.} 2020, \apj, 905, 23,
  \dodoi{10.3847/1538-4357/abca90}

\end{thebibliography}

\end{document}